\documentclass[modern]{aastex61}
\usepackage{amsmath}
\usepackage{color}
\renewcommand{\twocolumngrid}{} 

\newcommand{\foreign}[1]{\textsl{#1}}
\newcommand{\acronym}[1]{{\small{#1}}}
\newcommand{\project}[1]{\textsl{#1}}
\newcommand{\tgas}{\project{\acronym{TGAS}}}
\newcommand{\tmass}{\project{\acronym{2MASS}}}
\newcommand{\psone}{\project{\acronym{PS1}}}
\newcommand{\gaia}{\project{Gaia}}
\newcommand{\panstarrs}{\project{Pan\acronym{STARRS1}}}
\newcommand{\xd}{\acronym{XD}}
\newcommand{\cmd}{\acronym{CMD}}

\DeclareMathOperator*{\argmax}{arg\,max}
\newcommand{\given}{\,|\,}
\newcommand{\dd}{\mathrm{d}}
\newcommand{\true}{\mathrm{true}}
\newcommand{\corr}{\mathrm{c}}
\renewcommand{\vec}[1]{\boldsymbol{{#1}}}
\newcommand{\truth}{\vec{Y}}
\newcommand{\data}{\vec{y}}
\newcommand{\mean}{\vec{\mu}}
\newcommand{\mat}[1]{{\mathbf{{#1}}}}
\newcommand{\var}{\mat{V}}
\newcommand{\covar}{\mat{C}}

\shorttitle{improving gaia parallax precision}
\shortauthors{anderson et al.}
\begin{document}\sloppy\sloppypar\raggedbottom\frenchspacing 
\title{Improving \textsl{Gaia} parallax precision \\
       with a data-driven model of stars}

\author[0000-0001-5725-9329]{Lauren Anderson}
\affil{Center for Computational Astrophysics, Flatiron Institute, 162 Fifth Ave, New York, NY 10010, USA\hspace*{-1cm}}

\author[0000-0003-2866-9403]{David W. Hogg}
\affil{Center for Computational Astrophysics, Flatiron Institute, 162 Fifth Ave, New York, NY 10010, USA\hspace*{-1cm}}
\affil{Center for Cosmology and Particle Physics, Department of Physics, New York University, 726 Broadway, New York, NY 10003, USA}
\affil{Center for Data Science, New York University, 60 Fifth Ave, New York, NY 10011, USA}
\affil{Max-Planck-Institut f\"ur Astronomie, K\"onigstuhl 17, D-69117 Heidelberg}

\author[0000-0002-3962-9274]{Boris Leistedt}
\affil{Center for Cosmology and Particle Physics, Department of Physics, New York University, 726 Broadway, New York, NY 10003, USA}
\affil{NASA Einstein Fellow}

\author[0000-0003-0872-7098]{Adrian M. Price-Whelan}
\affil{Department of Astrophysical Sciences, Princeton University, 4 Ivy Lane, Princeton, NJ 08544, USA}

\author[0000-0001-6855-442X]{Jo Bovy}
\affil{Center for Computational Astrophysics, Flatiron Institute, 162 Fifth Ave, New York, NY 10010, USA\hspace*{-1cm}}
\affil{Department of Astronomy and Astrophysics, University of Toronto, 50 St. George Street, Toronto, ON M5S 3H4, Canada}
\affil{Alfred P. Sloan Fellow}

\begin{abstract}\noindent 
Converting a noisy parallax measurement into a posterior
belief over distance requires inference with a prior.
Usually this prior represents beliefs about the stellar density distribution of
the Milky Way.
However, multi-band photometry exists for a large fraction of the
\gaia\ \tgas\ Catalog and is incredibly informative about stellar distances.
Here we use \tmass\ colors for $1.4$ million \tgas\ stars to build a
noise-deconvolved empirical prior distribution for stars in color--magnitude
space.
This model contains no knowledge of stellar astrophysics or the
Milky Way, but is precise because it accurately generates a large number of noisy
parallax measurements under an assumption of stationarity; that is,
it is capable of combining the information from many stars.
We use the Extreme Deconvolution (\xd) algorithm---an Empirical Bayes approximation to a full
hierarchical model of the true parallax and photometry of every star---to
construct this prior.
The prior is combined with a \tgas\ likelihood to infer a precise photometric parallax estimate
and uncertainty (and full posterior) for every star.
Our parallax estimates are more precise than the \tgas\ catalog entries by a
median factor of 1.2 (14\% are more precise by a factor $>2$) and are more
precise than previous Bayesian distance estimates that use spatial priors.
We validate our parallax inferences using members of the Milky Way
star cluster M67, which is not visible as a cluster in the \tgas\ parallax
estimates, but appears as a cluster in our posterior parallax estimates.
Our results, including a parallax posterior pdf for each of 1.4
million \tgas\ stars, are available in companion electronic tables.
\end{abstract}


\section{Introduction}

The \gaia\ Mission (\citealt{gaia16}) will soon deliver more than a billion
stellar parallax measurements.
Only a small fraction (but large number) of these
measurements will be precise and purely astrometric.
\gaia\ will use astrometric parallaxes to determine the distances of the more
precise stars, and then calibrate spectrophotometric models (\citealt{apsis}).
These spectrophotometric models, along with \gaia's on-board
low-resolution $BP/RP$ spectrophotometry, can be used to provide
more precise parallax estimates for stars with poor astrometric parallax data.
The full stack required for these parallax inferences is complex.
It involves modeling the stars, as well as the dust, in the Milky Way,
and the response of the telescope itself.
It might be possible to eliminate some of this complexity using data driven models which don't rely on physical models.

Projects like \project{The Cannon} (\citealt{ness15}; \citealt{casey17}; \citealt{ho17})
and \project{Avast} (Bedell et al.\ in preparation),
explore the extent to which our predictive models of
stars could be purely data-driven or statistical.
That is, under what circumstances could the data themselves deliver
more precise or more accurate information than any theoretical or
physics-based model?
The answer to this question is extremely context-dependent: It depends
on what data are available, how much data are available, and what
questions are being asked.
In general, data-driven models contain fewer assumptions than
physics-driven models, but they are also usually less interpretable;
the model presented here has these same properties.
Yet, the \gaia\ data set is ideal for thinking about these kinds of purely statistical
models of stars.
The different stars observed by \gaia\ are measured at very different
signal-to-noise ratios,
and a data-driven model of the stellar color--magnitude distribution can capitalize
on the precise information coming from the high signal-to-noise stars
and deliver valuable information about the low signal-to-noise stars.
This information can then deliver precise parallax (or distance or absolute magnitude) inferences for all of the stars.

Here we present a demonstration-of-concept that shows that a
data-driven model of the \gaia\ data could in principle deliver very
precise parallax measurements, without any input of the physics or numerical models of stellar
evolution, interiors, or photospheres.
That is, this is a data-driven or purely empirical \emph{photometric
  parallax} method.
A photometric parallax is a parallax (or distance) estimate based on
the photometry (colors and apparent magnitude or magnitudes) of a
star. This idea was pioneered by \cite{juric08} using the Sloan Digital Sky Survey (\citealt{sdss}).
Fundamentally, photometric parallaxes capitalize on the fact that the
apparent magnitude of a star is a strong function of distance, and
that the absolute magnitude is a strong function of color (or surface
temperature) and evolutionary stage.

Most stellar models used to generate photometric
parallaxes have been \emph{physical} models, based on gravity, fluid mechanics,
radiative transfer, nuclear reactions, and atomic and molecular transitions.
Because there are small issues with each of these components, the physical models are inaccurate in detail, and
therefore produce parallax (and distance) estimates that are biased.
In addition, they build a long list of physical assumptions (about
nuclear reactions, convection, and thermal timescales, for example)
into the parallax estimates.
When using physical models, it is impossible to deliver photometric parallax estimates that involve minimal assumptions.

However, the use of physical models for distance estimation is not necessary.
It is possible to build a stellar photometric model from the data themselves,
because there are stars at a wide range of luminosities with useful parallax information, and this
range of stellar luminosities with useful parallax information will grow with future \gaia\ data releases.
One challenge is that different stars are observed at different levels of parallax precision,
so it requires relatively sophisticated technology to build this data-driven model
using all of the data available, fairly and responsibly. 

Here we build and use just such a data-driven model to infer more precise parallaxes for each star.
In particular, we use all of the \gaia\ \tgas\ data (\citealt{tgas};
these data were generated by the methodology of \citealt{michalik15}), that match to the
\tmass\ photometry (\citealt{skrutskie06}) and lie in the footprint of \panstarrs\ (\psone; \citealt{ps1})
to make use of the \cite{green15} dust map. We use this data to make a model of the
noise-free (or low-noise) color--magnitude diagram (\cmd) of stars.
We build the \cmd\ using an Empirical-Bayes approach known as Extreme Deconvolution (XD;
\citealt{bovy11}).
This method deconvolves heteroskedastic data to derive an
estimate of the noise-free or low-noise distribution that \emph{would
  have} been observed with far better data.
This method has been used in astrophysics
previously to model the velocity distribution of stars in the Solar
Neighborhood (\citealt{hogg05}; \citealt{bovy09}) and to
perform quasar target selection
in \project{Sloan Digital Sky Survey} data (\citealt{xdqso}; \citealt{xdqsoz}).
Its advantage over other deconvolution methods is that it takes as input,
and handles in a principled way, heteroskedastic data.
Its disadvantage, for the present purposes, is that it requires
or implies a strictly Gaussian noise model.
As we show below, this requires us to transform the \cmd\ to a space in which
the noise is approximately Gaussian.

We then use the \xd\ output---the deconvolved \cmd---as
a prior for use in inference of individual stellar parallaxes.
These inferences, one per star, provide much more precise parallax, distance,
or absolute magnitude estimates than we get from each star's primary
\tgas\ parallax measurement alone.
That is, we are using the \xd\ model to de-noise the \gaia\ parallax
measurements.
Technically, since \xd\ produces a maximum-marginalized-likelihood estimator
of the deconvolved distribution,
its use as a prior is technically using the data twice and is therefore only approximately valid.
However, since the data set is large, the \xd\ approximation is not bad; it is
sometimes known as the ``empirical Bayes'' method, and is well studied.\footnote{For a primer on ``empirical Bayes'', see \url{https://en.wikipedia.org/wiki/Empirical_Bayes_method}.}

A probabilistically valid approach that wouldn't re-use the data is to perform
a full hierarchical Bayesian inference.
Exploration of that possibility, and its computational tractability,
is among our long-term goals and motivations.
Along those lines, this paper can be seen as a companion to a related
project \citep{leistedtHogg2017}, in which we use a slightly less appropriate
model for the \cmd, but in which we take a much more principled
approach to the inference.

Although not a fully Bayesian hierarchical inference, the methodology presented in this paper makes exceedingly weak assumptions
about the Milky Way, and similarly weak assumptions about the properties of
stars.
To date, there has been some discussion in the literature
about how to use a parallax responsibly to infer a distance (\citealt{astraatmadja16a}; \citealt{astraatmadja16b}).
These papers use a \gaia\ measurement to construct a parallax likelihood,
and combine this with a distance prior.
While this is sensible and correct, all present methods proposed along
these lines build in informative (and known-to-be-wrong) assumptions
about the line-of-sight distance distribution to \gaia\ stars.
That is, they build in strong and informative assumptions about the Milky
Way.
The strongest new assumption made in this work is that for every star
in \tgas, there are other, photometrically (and bolometrically)
similar stars.
That is a much weaker assumption than has been made in most other Bayesian
distance estimations.

This project is fundamentally a demonstration of concept:
We are only using a subset of the ``small'' (relative to the expected final \gaia\ data set)
\tgas\ Catalog.
We are performing only an approximation to full Bayesian hierarchical inference.
We have to use photometry that is ground-based, rather than the full-precision,
space-based photometry that the \gaia\ Mission will deliver.
However, as we show below, we get very good results in terms of parallax precision.
It is promising that we will be able to obtain even better results
using later \gaia\ data releases, and eventually infer distances for $>10^9$
\gaia\ stars using photometric parallax methodologies but without any
commitment to---or even use of---physical models of stars or the Milky Way.

\section{Why does this work?}
\begin{figure}
\centering
  \includegraphics[width=\textwidth]{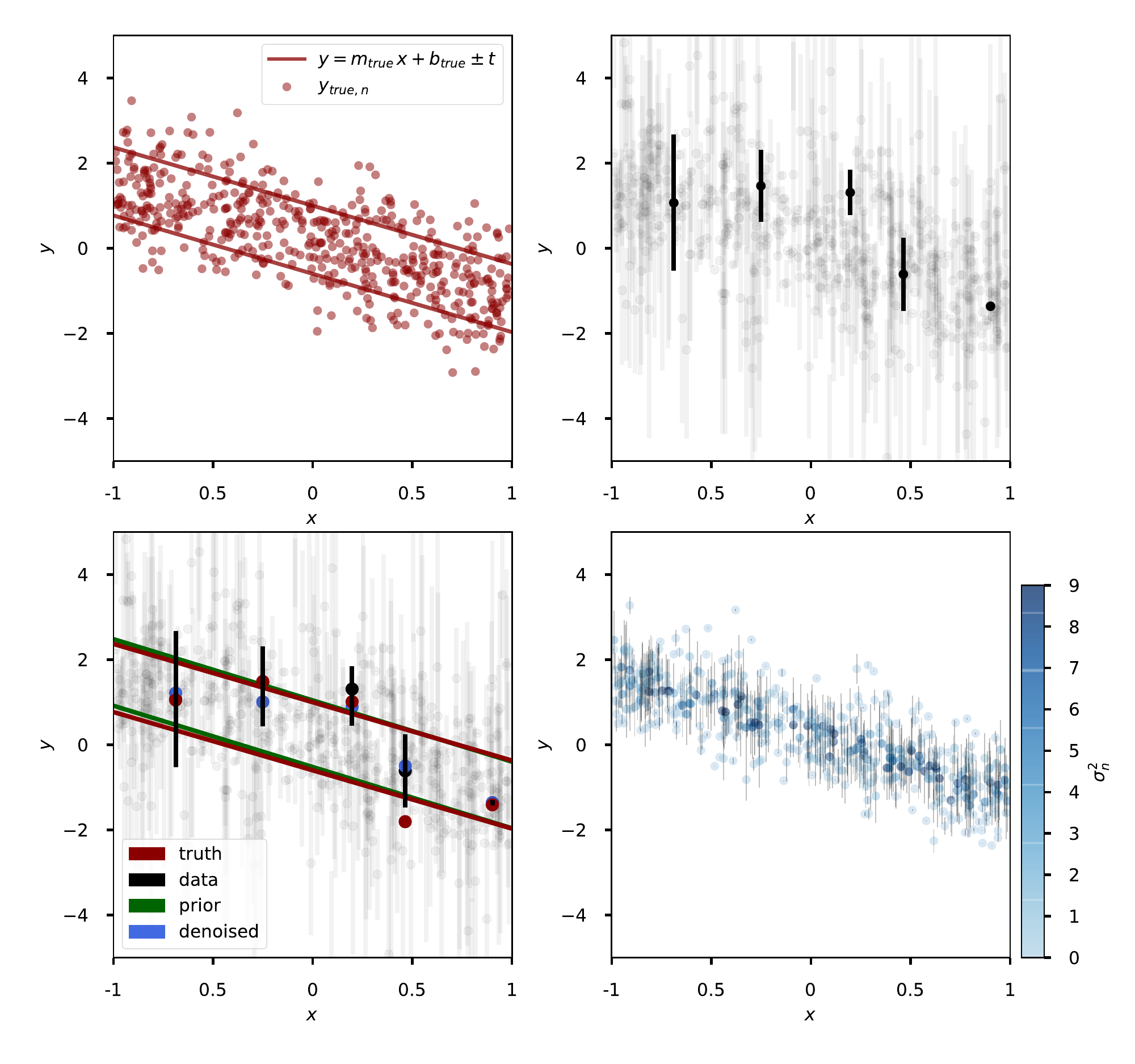}
\caption{ {\bf Why does this work}: An example of generating posterior beliefs of true values from noisy data by using the noise-deconvolved distribution of data as a prior. The upper-left panel visualizes the simplified toy distribution. The red lines are the $1\sigma$ contours of the distribution, and the red points are 1024 samples from the distribution. These same $1\sigma$ contours are drawn in the bottom left panel. In the upper-right and lower-left panel, the grey points are these same 1024 samples of the distribution but with some measurement uncertainty added to them. The error bars show the $1\sigma$ measurement uncertainties. The black points are 5 randomly chosen data to highlight the method. In the lower-left panel, the green lines represent the noise deconvolved, best estimate for the underlying distribution. The blue points represent the expectation value of the posterior belief of the true values of the noisy black points using the noise deconcolved distribution as the prior. The red points represent the actual true values. The lower-right panel shows the expectation value, and $1\sigma$ uncertainties, of the posterior belief of the true value for all the points. They are colored by the measurement uncertainty, showing that data with larger measured variance are more influenced by the prior and lie towards the center of the noise deconvolved distribution, or prior. In contrast, more certain measurements tend to lie closer to their measured value.}
\label{fig:toy}
\end{figure}

In this paper we use the noise-deconvolved \cmd\ as a prior to infer more precise parallax probability distribution functions (PDFs) from \gaia\ measurements.
Understanding the validity of this method can be challenging because we are using the data both for the prior and the likelihood in our inference.
To ease understanding, here we demonstrate the general methodology using a
simpler toy model with simulated data.
Instead of inferring parallax PDFs using the 2D distribution of the \cmd\ as our prior, for this simpler model we will infer posterior PDFs of the value $y$.
We perform this simpler inference using a prior that is also a 2D distribution: a 1D Gaussian in the $y$ direction, with a running mean $\mu = m\,x + b$ and some thickness $t$.
Here the thickness of the toy distribution mimics the true thickness, or intrinsic scatter, of the \cmd\ due to age, mass, and metallicity.
The $1\sigma$ contours as well as (noise free) samples from the toy model distribution are shown in \figurename~\ref{fig:toy}, upper-left panel in red.
Similarly, noisy samples from this distribution are shown in the upper-right panel in grey and black.
Using a representation of the true distribution (which we learn from the noisy data) as our prior, we can infer some posterior belief over $y_{\true,n}$, a true sample from the distribution, from $y_n$, a sample from the noisy distribution.

Setting up the problem in more detail, we have the true samples from the underlying distribution and the noisy samples
\begin{eqnarray}
y_{\true, n} \, &=& \, m_{\true}\,x_n + b_{\true} + \Delta_n \\
y_n \, &=& \, y_{\true,n} + \epsilon_n \quad,
\label{eq:ytrue}
\end{eqnarray}
where $\Delta_n$ is drawn from the Gaussian function $\mathcal{N}(\Delta_n \given 0, t^2)$---which accounts for the intrinsic width---and $\epsilon_n$ is drawn from the Gaussian function $\mathcal{N}(\epsilon_n \given 0, \sigma_n^2)$---the noise distribution---so $\sigma_n$ is the measurement uncertainty.
These represent true and noisy draws from a distribution that is a straight line with some thickness $t$, and are visualized in the upper panels of \figurename~\ref{fig:toy}.
The upper-left shows 1024 realizations of the true values, and the upper-right, as well as lower-left, show 1024 realizations of the noisy values $y_n$ and their associated measurement uncertainties $\sigma_n$ as the grey points; we highlight 5 random points in black.

To infer a posterior PDF of the true $y$ value for each measured $y$ value we need a prior. We use an estimate of the underlying true distribution that the noisy data $y_n$ are drawn from, which we build from the noisy $y_n$ data themselves.
We learn the noise-deconvolved function using marginalized, maximum-likelihood estimation (MMLE), which marginalizes over the true data values and maximizes the likelihood of the data. Here our likelihood for each individual datum is
\begin{eqnarray}
p(y_n \given m, b, t^2, \sigma_n^2) \ &=& \ \int \mathcal{N}(y_n \given y_{\true,n}, \sigma_n^2) \, \mathcal{N}(y_{\true,n} \given m\,x_n + b, t^2 ) \, \dd y_{\true, n} \\
&=&  \ \mathcal{N}(y_n \given m\,x_n + b, t^2 + \sigma_n^2) \quad,
\label{eq:toyLike}
\end{eqnarray}
where we have marginalized over the true sample $y_{\true, n}$. The resulting likelihood is a Gaussian because the noise model and our assumptions about the intrinsic width are both Gaussian. The likelihood of the full data set $\{y_n\}$ is then the product of the individual likelihoods
\begin{equation}
p(\{y_n\} \given m, b, t^2, \sigma_n^2) = \prod_n\, p(y_n \given m, b, t^2, \sigma_n^2) \quad .
\label{eq:toyLikeFull}
\end{equation}
Maximizing this joint likelihood of all the data is the same as minimizing
\begin{equation}
\chi^2 = \sum_n \frac{[y_n - [m\,x_n + b]]^2}{[\sigma_n^2 + t_n^2]} + \sum_n\,\ln(2\pi\,[\sigma_n^2 + t^2])
\quad .
\label{eq:chisq}
\end{equation}

The MMLE gives us the best fit parameters $\hat{m}$, $\hat{b}$, and $\hat{t}$. In \figurename~\ref{fig:toy}, lower-left panel, the true functions $y_{\true} = m_{\true}\,x + b_{\true} \pm t_{\true}$ are shown as the red lines, and the best fit MMLE functions $\hat{y}_{\true} = \hat{m}\, x + \hat{b} \pm \hat{t}$ are shown as the green lines. Now that we have an estimate of the true underlying distribution $\hat{y}_{\true}$, we use this as a prior to infer a posterior belief of the true values for each datum $p(y_{\true,n} \given y_n, \tilde{\sigma}_n^2)$.

Here is a good point to step back and acknowledge that our method is a
bit strange. We are using the data to determine the
noise-deconvolved distribution (an estimate of the true
distribution), and then using that noise-deconvolved distribution to
make inferences about the data. We are technically using the data
twice! To be fully, technically correct, we should \emph{either} learn the true
distribution while leaving one datum out, the one datum we want to
infer the true y value for, or else perform a full hierarchical Bayesian inference,
in which we treat the \cmd\ parameters probabilistically along with everything else.
With large data sets, the MMLE or Empirical Bayes approximation is safe,
in some sense, because no individual data point makes a large difference to
the inference.

Returning to the inference:
We use Bayes' theorem to turn the likelihood of the noisy data into a posterior belief over the true $y$ value:
\begin{eqnarray}
p(y_{\true,n} \given y_n, \tilde{\sigma}_n) &=& p(y_n \given y_{\true,n}, \sigma_n)\,p(y_{\true,n}) \\
\mathrm{where}\;p(y_n \given y_{\true,n}, \sigma_n) &=& \mathcal{N}(y_n \given y_{\true,n}, \sigma_n) \\
\mathrm{and}\;p(y_{\true,n}) &=& \mathcal{N}(y_{\true,n} \given \hat{m}\, x_n + \hat{b}, \hat{t})
\label{eq:toyBayes}
\end{eqnarray}
In \figurename~\ref{fig:toy} lower-left panel, the blue points represent the posterior beliefs (visualized as the expectation value and $1\sigma$ uncertainties) of the true values $y_{\true,n}$ for the 5 highlighted (black) points from the upper-right panel. The red points are the associated true $y$ values.
In the lower-right panel of \figurename~\ref{fig:toy}, the blue points represent the posterior beliefs, again visualized as the expectation value and $1\sigma$ uncertainty, for all 1024 $y_n$.
When the true values are inferred in this way, the associated variance $\tilde{\sigma}_n^2$ is the convolution of the true distribution variance, $t^2$, and the measurement variance, $\sigma_n^2$.
The points are colored by their measurement variance $\sigma_n^2$, which shows that noisier measurements are more influenced by the prior and get pulled closer to the center of the MMLE distribution, our prior. In contrast, more certain measurements remain closer to their measured values, $y_n$. We now have better estimates of each true $y$ value by inferring each true $y$ value from the measured $y$ value using an estimate of the true distribution the measured $y$ value was drawn from as the prior.

In detail, the posterior points shown in the lower-right panel of
Figure~\ref{fig:toy} look more concentrated (vertically) than the true
values shown in the upper-left panel.
This is because the points in the lower-right (posterior) panel show
not typical values or typical draws from the posterior; they show the
modes (and, with error bars, the variances) of the posterior PDFs.
When a measured value has a large variance, the posterior ends
up close to the prior, which has a mode at the center line; the
posterior modes therefore concentrate more towards the center line
than any representative sampling.
A plot of posterior samplings (as opposed to a plot of posterior modes) would
show a scatter like that in the upper-left (truth) panel.

Bringing the toy model back to our original problem:
We can get better estimates of the true parallax for each star by inferring the true parallax from the measured parallax and using the noise-deconvolved \cmd, an estimate of the true \cmd, as our prior.
For the remainder of the paper, we will also maintain the color scheme presented here (where possible): truth will be red, data will be black, prior information will be green, and posterior information will be blue.

\section{Assumptions and methods}\label{sec:method}

To infer the true parallax for a star using the noise-deconvolved \cmd\ as
a prior, we make the following assumptions. The method we present is correct
and justifiable under these assumptions, each of which is questionable
in its own right. We return to criticize these assumptions---and the
method that flows from them---in the discussion section below.
\begin{description}
\item[stationarity] We make an assumption of \emph{stationarity}; that is, that the
  stars measured at lower signal-to-noise have analogs measured at
  higher signal-to-noise.
  In detail, this assumption is wrong. For example, more distant stars will have lower signal to noise ratio measurements
  and will also be different than local
  disk stars in age and metallicity. However, the stationarity assumption is fairly weak:
  the requirement is that, locally in the \cmd, there is
  support among the high signal-to-noise stars for the types
  of stars also seen at low signal-to-noise.
  If this requirement is met, the results will not be
  strongly biased. Although this is a weak assumption, it is deep and fundamental to this
  work.
\item[selection] We have no model for the selection function of \tgas\ (but see \cite{bovy17}),
  nor selection volumes for any kinds of stars.
  For this reason, we are building a model of \emph{the contents of \tgas}, not the
  properties of the stars in the Milky Way (nor any volume-limited component or region
  thereof). For this reason, our \cmd\ prior might not be appropriate to use with other surveys.
\item[big data] We assume that we have large numbers of stars, large enough that
  empirical-Bayes, or maximum-marginalized-likelihood, is a safe
  approximation to full Bayesian inference. This assumption will be
  least true in the least-populated parts of the \cmd. In the extreme
  case, if a portion of the \cmd\ has only one (statistically distinct)
  exemplar---for example, a white dwarf---inference for these cases will be biased.
\item[noise model] We assume that the \tgas\ parallax uncertainties dominate the
  noise in any estimates of absolute magnitude. We further assume that
  the parallax and color uncertainties are Gaussian in form with
  correctly known variances. In detail, we assume that the parallax uncertainty estimates,
  (the \texttt{parallax{\_}error} entry in the \tgas\ catalog), are correct.
  Here we are making use of the fact that when
  de-noising data, under estimating measurement noise is conservative, and leads to under-deconvolution. In other words, the features of the inferred \cmd\ will typically be broader and more conservatively estimated than if we over-estimated the noise.
\item[dust] We treat the \psone-based three-dimensional dust maps (\citealt{green15})
  as correct in their median dust estimates, and their
  effects on color and magnitude. As we optimize the empirical Bayes model, we iteratively update each star's
  dust correction at its updated, inferred distance. Our assumption is that these corrections are
  good enough, and uncertainties in these are not dominant, for neither
  color nor absolute magnitude.
\item[mixture of Gaussians] We assume that the \cmd\ can be represented by a mixture of
  Gaussians in a particular transformed space. This mixture we fix
  (with only heuristic analysis) at 128 Gaussians.
\item[no physics] We make no use of stellar models; our only assumptions about
  stellar physics are generic and implicit (for example, that the
  \cmd\ is somehow smooth and compact).
\end{description}

\subsection{Setting up the Inference}

Under these assumptions, we would like to infer the true parallax, $\varpi_{\true}$, for each star given its measured parallax, $\varpi$, and associated uncertainty, $\sigma_{\varpi}$, from \tgas.
A first application of Bayes theorem leads to
\begin{equation}
p(\varpi_{\true} \given \varpi, \sigma^2_{\varpi}) = \frac{1}{Z}\,p(\varpi \given \varpi_{\true}, \sigma^2_{\varpi}) \, p(\varpi_{\true}) \quad ,
\label{eq:bayes}
\end{equation}
where $p(\varpi \given \varpi_{\true}, \sigma^2_{\varpi})$ is the likelihood of the observed parallax, $p(\varpi_{\true})$ is a prior PDF for the true
parallax, and $Z$ is a normalization constant (the model evidence, or marginalized likelihood).
The strength of our inference lies in how we generate our prior.
Instead of assuming a functional form for the spatial distribution of stars in the Milky Way (\citealt{astraatmadja16b}), or relying on stellar models (\citealt{gaia16}), we build a noise-deconvolved estimate of the \cmd\ from the observed colors and absolute magnitudes of our full set of stars.
This is because supplementary to the trigonometric measure of distance, the photometry of stars is also informative of their distances.
The intrinsic temperatures and luminosities of stars tightly correlate, leading to the observed colors and absolute magnitudes of stars also tightly correlating, usually visualized as the \cmd.
The tight correlation of the \cmd\ benefits our inference because the luminosity or absolute magnitude of a star $M$, when compared with its brightness or apparent magnitude $m$, contains distance information $d$ via the fundamental relation $m=M+5\log(d/10\, \mathrm{pc})$.
We include this in the inference by using the \cmd\ as part of our prior for $\varpi_{\true}$.
Specifically, we use the \cmd\ generated from the data themselves, as detailed in the next section.

Equation~\ref{eq:bayes} does not fully specify this procedure.
Because our prior comes from the \cmd, our inference will include not only the parallax from \tgas, but also the photometry of the stars. Specifically, we use the $J$ and $K_s$ band photometry from \tmass, as well as the parallax from \tgas. Our true, latent values are the $(J-K_s)^{\true}_n$ color and the $J$ band absolute magnitude $M^{\true}_{J,n}$ for each star, which we encapsulate in the true vector $\truth$. The observed data is the dust corrected color $(J-K_s)^{\corr}_n$, the \tgas\ parallax $\varpi_n$ in units of $[mas]$, and the dust corrected $J$ band magnitude $J^{\corr}_n$, which we encapsulate in the data vector $\data$.

We now describe a singularity of our method, which allows us to conservatively deconvolve the data into a \cmd\ prior without requiring a full hierarchical inference framework (which was explored in \citealt{leistedtHogg2017}).
Instead of transforming the parallax and apparent magnitude into an absolute magnitude, the data vector $\data$ contains a transformation of the dust corrected absolute magnitude
\begin{equation}
y_0 = \varpi\,10^{\frac{1}{5}\,J^{\corr}_n} = 10^{\frac{1}{5}\,M^{\corr}_{J,n} + 2}
\label{eq:transform}
\end{equation}
where, for this expression, it is important to remember that $\varpi$ is assumed to be in units of $[mas]$.
The transformation of the (dust corrected) absolute magnitude is required to keep the uncertainties Gaussian, and Gaussian uncertainties are required to use \xd, see below.
Our prior assumptions $I$ are the uncertainties in each of these measurements, as well as the 3D dust map. So our full terminology is
\begin{equation}
\begin{aligned}
\mathrm{true \, vector} \quad &\truth_n = [10^{0.2\,M^{\true}_{J,n} + 2}, (J-K_s)^{\true}_n] \\
\mathrm{data \, vector} \quad &\data_n = [\varpi_n\,10^{0.2J^{\corr}_n}, (J-K_s)^{\corr}_n] \;\hspace*{15mm}  n \in \mathrm{stars} \\
\mathrm{with \; apparent \; magnitude} \quad &J^{\corr}_n = J_n - Q_J\,E(B-V)_n \\
\mathrm{and \; color} \quad &(J-K_s)^{\corr}_n = J_n - K_{s,n} - Q_{JK}\,E(B-V)_n \\
\mathrm{under\ the\ assumptions} \quad &I_n = \left\{\left\{\sigma^2_{\varpi, n}\right\}, \left\{\sigma^2_{J,n}\right\}, \left\{\sigma^2_{K,n}\right\}, \mathrm{Dust \, Map}\right\}
\end{aligned}
\label{eq:data}
\end{equation}
where $E(B-V)_n$ is the output of the dust model of \cite{green15}, and $Q_{\lambda}$ is the correction factor for a photometric band assuming some reddening law.

We assume that the parallax likelihood is Gaussian, as well as the $(J-K_s)^{\corr}_n$ color likelihood. So our full likelihood for the $n$th star is
\begin{equation}
p(\data_n \given \truth_n, I_n) = \mathcal{N}(\data_n \given \truth_n, \covar_n)
\end{equation}
with \\
\[
\covar_n = \begin{bmatrix}
\left(\sigma_{\varpi}\,10^{0.2\,J_n^{\corr}}\right)^2 & 0 \\
0 & \sigma_{J,n}^2 + \sigma_{K,n}^2
\end{bmatrix}
\]
where $\mathcal{N}(\mean, \mat{\Sigma})$ represents a normal distribution with
mean vector $\mean$ and covariance matrix $\mat{\Sigma}$.
Here we have assumed that the parallax uncertainty is significantly larger than the $J$-band magnitude uncertainty and is therefore the dominant uncertainty in the transformed absolute magnitude. In detail, we take $J_n^{\corr}$ as a point estimate and only propagate the parallax uncertainty.

We can now write the multidimensional posterior expression analogous to Equation~\ref{eq:bayes},
\begin{equation}
p(\truth_n \given \data_n, I_n) \propto p(\data_n \given \truth_n, I_n) \, p(\truth_n) \label{eq:posterior}
\end{equation}
where $p(\data_n \given \truth_n, I_n)$ is the likelihood of the data vector, and $p(\truth_n)$ is our prior.
Before we describe how this can be turned into a posterior belief on the true parallax, we turn our attention to the prior, which will be generated from the data themselves. For this, we use XD, which takes advantage of the Gaussian likelihood.

\subsection{Generating the Prior}

To generate the empirical prior $p(\truth_n)$,
we fit our data in the transformed \cmd\ space (see previous sub-section) using \xd,
which deconvolves the data $\{ \truth_n \}$ with heterogeneous,
Gaussian noise variances $\{ \covar_n\}$. \xd\ generates an estimate of the underlying distribution from which the uncertain data were drawn, modeled as a mixture of $K$ Gaussians:
\begin{equation}
	p\bigl(\ \truth \given  \left\{A_k, \mean_k, \var_k\right\}_{k=1}^K \bigr) = \sum_{k=1}^K A_k \ \mathcal{N}(\truth \given \mean_k, \var_k )
\end{equation}
where $\mean_k$ and $\var_k$ are a two dimensional vector and matrix, respectively, which correspond to the location and covariance of the $k$th component (a Gaussian in our transformed \cmd\ space). The relative amplitudes of the components are given by the scalars $\{ A_k \}$.

Given the data $\{ \truth_n, \covar_n\}$, \xd\ maximizes the marginalized likelihood of the data, marginalizing over the true data. Assuming $\alpha = \left\{A_k, \mean_k, \var_k\right\}_{k=1}^K$, it finds the coefficients
\begin{eqnarray}
\hat{\alpha} &=& \argmax_{\alpha} \, \prod_n \, p(\data_n \given \alpha) \nonumber\\
             &=& \argmax_{\alpha} \, \prod_n \, \int p(\data_n \given \truth_n) \, p(\truth_n \given \alpha)\, \dd\truth_n
\label{eq:xdmml}
\end{eqnarray}
which is made analytically tractable with the Gaussian likelihood and mixture-of-Gaussians prior \citep{bovy11}.

DWH: write something about how our prior is a prior on distance

\subsection{Back to the Inference}

To infer $\varpi_{\true}$ once the \cmd\ prior is constructed, we first do the inference in the 2D space of the \cmd\ prior $p(\truth \given \alpha)$, using Equation \ref{eq:posterior}. We then project the 2D posterior into the 1D $\varpi_{\true}$ space by multiply by the Jacobian to calculate the posterior PDF over $\varpi_{\true}$

\begin{eqnarray}
p(\varpi_{\true} \given \data, \hat{\alpha}) \, &\propto& \, \left|\frac{\dd\truth}{\dd\varpi}\right| \, p(\truth \given \data, \hat{\alpha}) \, f(\varpi_{\true}) \\
\mathrm{where} \; f(\varpi_{\true}) &=& \begin{cases}
              1 \quad \mathrm{if} \quad -2 < \mathrm{log}_{10} \varpi_{\true} < 2\\
              0 \quad \mathrm{otherwise}
              \end{cases}
\label{eq:parallaxPost}
\end{eqnarray}
with $f(\varpi_{\true})$ being a window function to insure $\varpi_{\true}$ is positive and to put $\varpi_{\true}$ on a similar grid for all the posteriors.

\section{Data and results}
\begin{figure}
\centering
  \includegraphics[width=\textwidth]{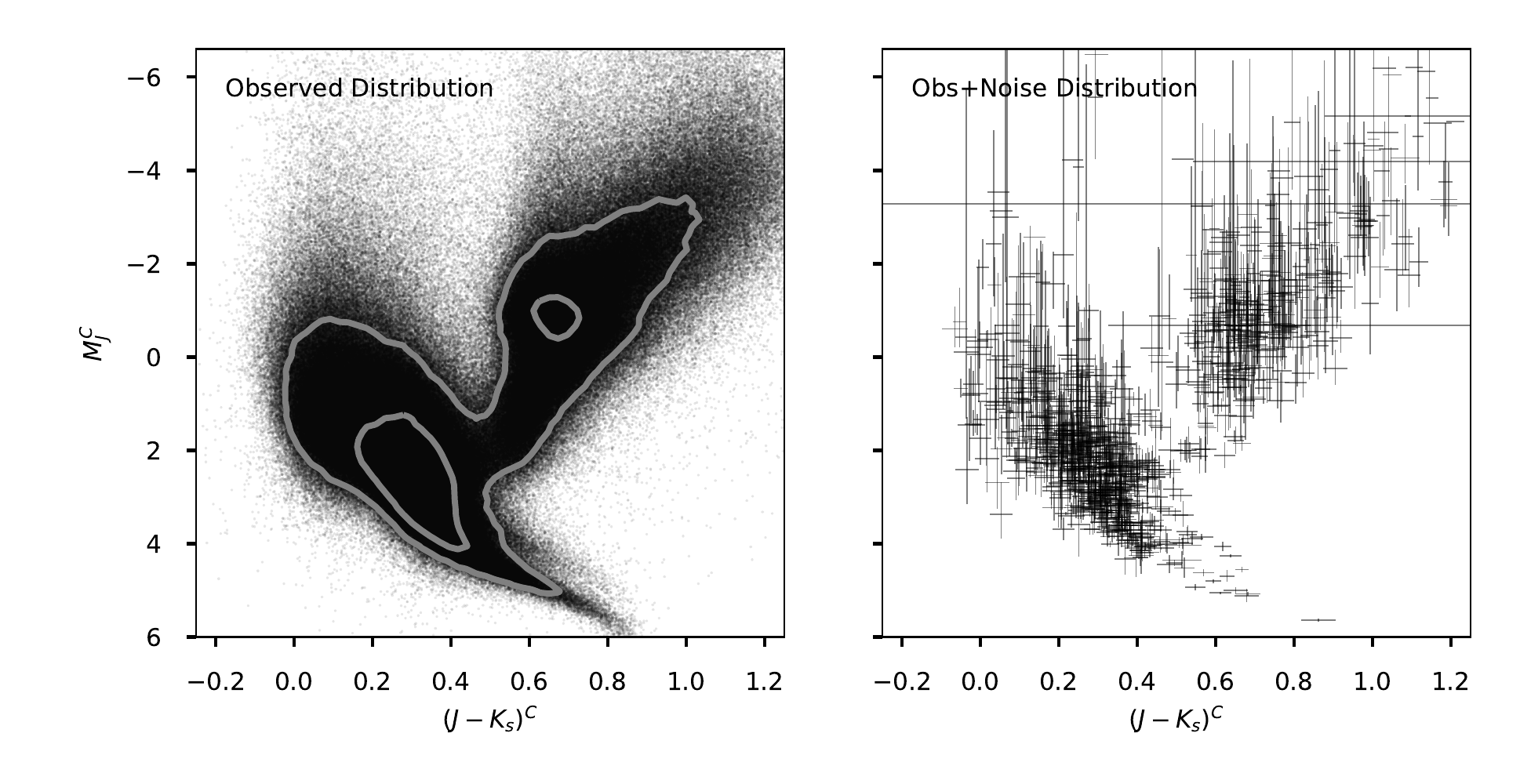}
\caption{The observed \cmd: To visualize the full dataset, in the left panel we use the point estimates of the \tmass\ $J-K_s$ color, the $J$ band apparent magnitude, and the \tgas\ parallax. The grey lines represent the $1\sigma$ and $2\sigma$ contours for the distribution. To give a sense of the uncertainties, in the right panel we subsample the dataset and include the $1\sigma$ uncertainties. The uncertainties are dominated by the parallax noise, with many stars diverging to infinitely far away and therefore very intrinsically bright.}
\label{fig:data}
\end{figure}

We use stars crossmatched in \tgas\ and \tmass.
The match was done using a nearest-neighbor algorithm with a search radius of $4$ arcsec\footnote{\url{http://portal.nersc.gov/project/cosmo/temp/dstn/gaia/tgas-matched-2mass.fits.gz}}.
We also required that the stars lie within the
observing footprint of \psone\ to access the \cite{green15} 3D dust model.
We require that the photometry have real values, and nonzero, real errors, and
remove a small selection of \tmass\ stars that have zero color and zero $J$-band
magnitude.
The full data set is visualized in the \cmd\ in \figurename~\ref{fig:data}.
The left panel shows the point estimates of the color and absolute magnitude,
using the point estimate of the parallax from \tgas. The $1\sigma$ and
$2\sigma$ contours of the distribution are shown in grey.
The right panel shows a subset of the data with the associated error bar for each star.
The uncertainties in the colors are fairly well behaved, but the large
uncertainties in some absolute magnitudes are due to the large uncertainties in
the parallax measurements.

\subsection{Dust}

To generate the prior and evaluate the likelihood, we need to
correct the \tmass\ photometry for dust extinction.
With the emergence of 3D dust maps, it is now possible to estimate dust
corrections to stars within the Milky Way. The challenge is getting a
proper estimate of the distance to the star before inferring
$\varpi_{\true}$, since we need a dust correction prior to doing our
inference. Dust corrections are particularly non trivial for stars
with poor parallax signal to noise; these stars have
a likelihood that is consistent with infinite distance and can
therefore have severe dust corrections. This is most obvious in the
giant stars, which tend to have the lowest signal to noise, and can
therefore have dust corrections $> 2$~mag. Although the dust
model is probabilistic, as well as our distance inference, to move
forward, we break from a fully Bayesian framework in which we might
infer the distance and the dust simultaneously (more about this in the
discussion). Instead, we correct for dust using a point estimate from
the 3D dust map \citep{green15}. We determine this dust correction by
iteratively inferring, and then sampling, our posterior PDF over
$\varpi_{\true}$. We first generate our prior using the observed,
attenuated photometry. Using this raw prior, we infer more precise
parallaxes to all the stars in our sample. We use this more
precise parallax posterior to get a measurement of dust in the 3D dust
map for each star. We then iterate this process ten times, however
the dust values seem to converge after a few iterations. The
iterative process allows for slight variations in distance and
corresponding dust to become more consistent with the \cmd.

\begin{figure}
\centering
  \includegraphics[width=\textwidth]{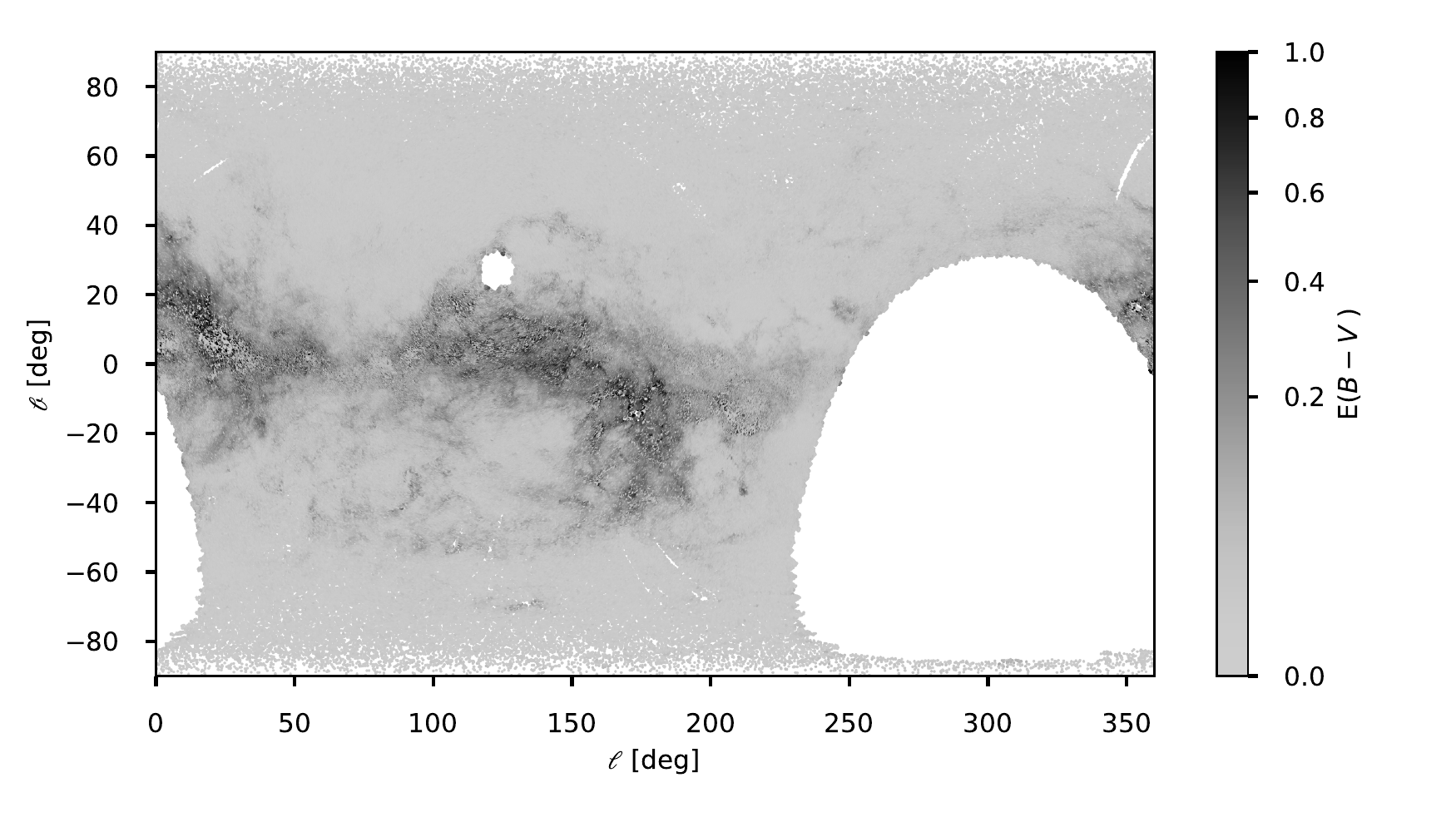}
\caption{The converged dust values at the 5\% distance quantile in Galactic coordinates. This shows both the footprint of our analysis as well as the adopted extinction values. Each point represents a single star in our cross-matched catalog. The large missing areas are due to the \psone\ footprint ($>-30~{\rm deg}$ in declination)}
\label{fig:dust}
\end{figure}

In detail, we take the $5\%$
quantile of each parallax posterior (the closest part) and query the 3D dust map at that distance.
We take the $5\%$ quantile---as apposed to the median---to do a minimal dust
correction because over-correcting for dust can push the likelihood
function into an unphysical, extremely blue part of the \cmd\ and thus bias the
parallax inference. At the 3D position of each star, we sample the \texttt{mode=median} value of the probabilistic dust map.
We apply each dust correction to our
\tmass\ photometry using Equation~\ref{eq:data}, where $Q_{\lambda}
= [0.709, 0.302]$ for bands $[J, K_s]$ respectively (see Table 6 in
\citealt{schlafly11}).
The dust values we obtain in our final
iteration, and which we apply to the photometry in our inference, are visualized in
galactic coordinates in \figurename~\ref{fig:dust}.
After applying the dust correction, we do not update the uncertainties, which is
technically wrong and under-represents the true uncertainties; this leads to a less severe deconvolution, as mentioned previously.
In other words, the features of our inferred
  \cmd\ are conservatively broad.\footnote{This can be understood by considering the following procedure: if one incrementally increased the noise levels (leaving the data unchanged; just increasing the declared uncertainties), the deconvolved distribution would be increasingly narrower, since it targets the distribution that would generate the noiseless observations, as well as the observed data once noise is added. Thus, under-estimating the uncertainties leads to a conservative under-deconvolution.}
We could add the covariance of
the dust to our likelihoods, but this is beyond the scope of this
demonstration of concept.

\subsection{The empirical prior}
\begin{figure}
\centering
  \includegraphics[width=\textwidth]{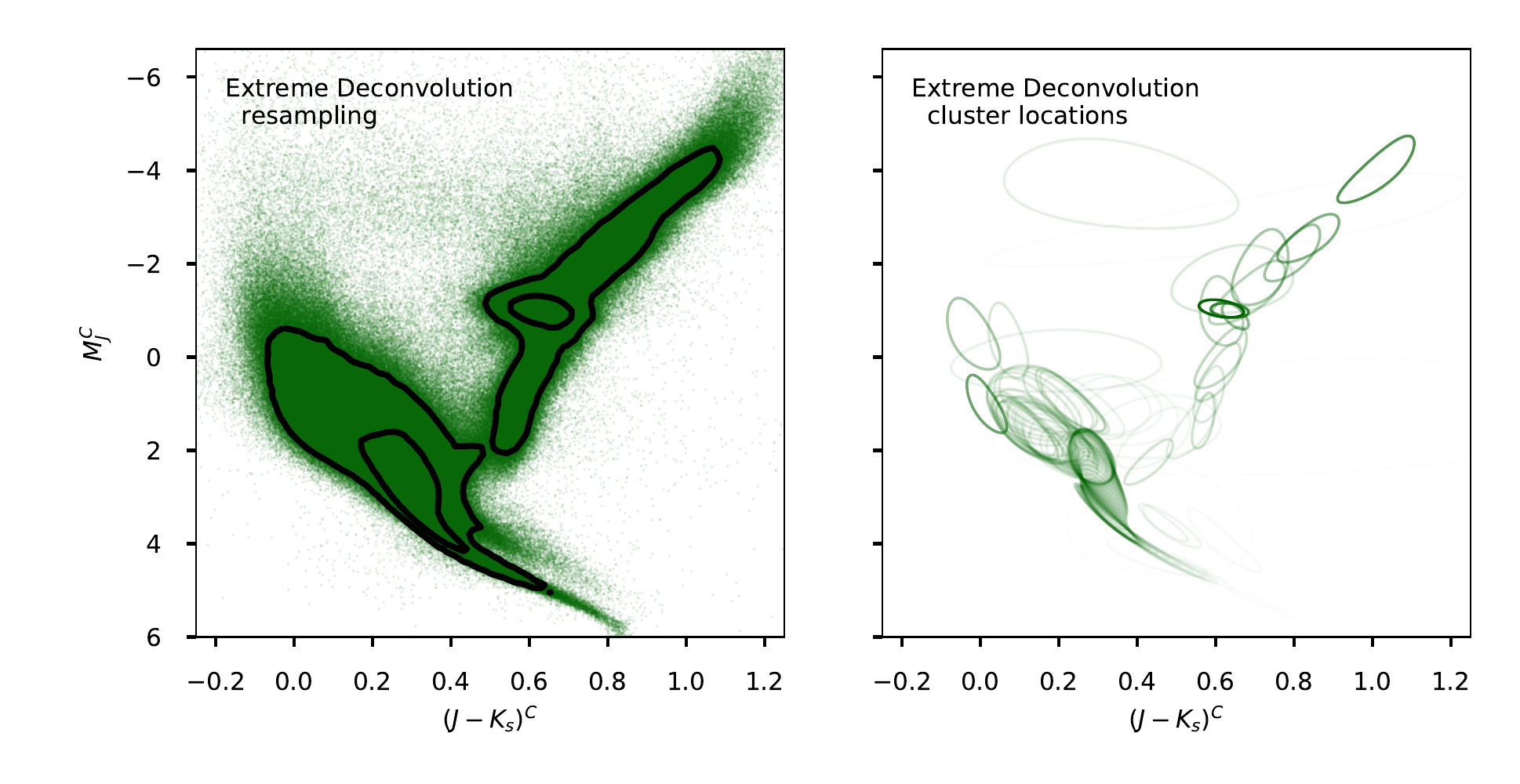}
\caption{The \cmd\ prior, modeled as a Gaussian mixture, inferred by
  running XD on all of the data. The left panel shows a sampling of
  the prior, with the black lines showing the $1\sigma$ and $2\sigma$
  contours. The right panel shows the 1-$\sigma$ contours of the
  individual components. The latter are Gaussian in the transformed
  magnitude space to make the XD inference tractable. Thus, they
  appear as slightly deformed ellipses in color--magnitude space.}
\label{fig:prior}
\end{figure}

Using \xd\ and the MMLE method described in the methods section, we generate the
prior, shown in \figurename~\ref{fig:prior}.
The left panel shows 1.4 million samples from the prior distribution, and the
right panel shows $1\sigma$ contours for each component in the underlying
Gaussian mixture model.
Compared to the raw data in \figurename~\ref{fig:data}, the deconvolved
color--magnitude diagram for the \gaia\ + \tmass\ stars is tighter, as expected.
We'll come back to more features in the data below, especially those seen in the posterior distribution.

\subsection{Shrinkage of parallax uncertainties}

\begin{figure}
\centering
  \includegraphics[width=\textwidth]{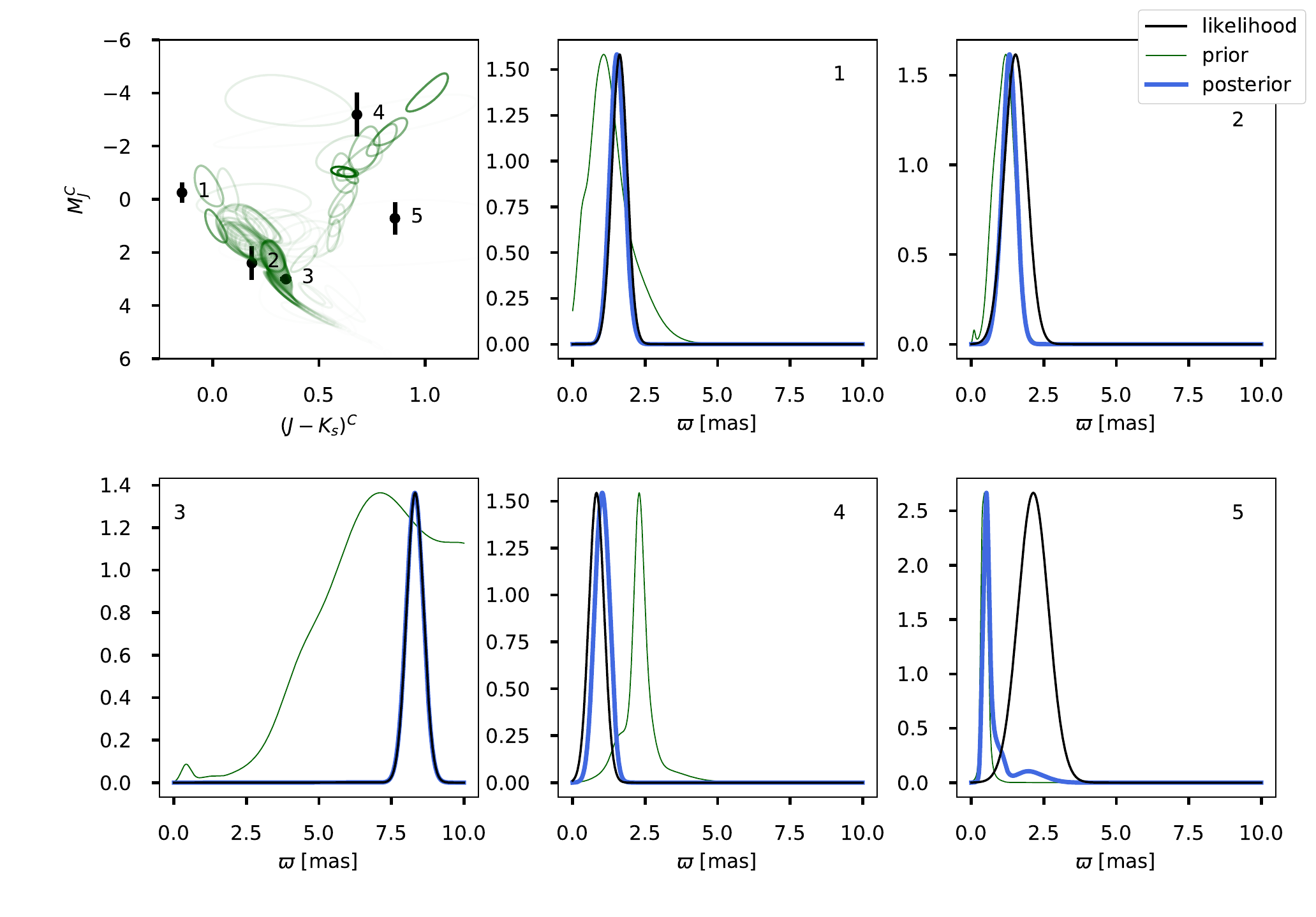}
\caption{Five example posterior PDFs over parallax using our method.
The upper left panel shows the position of each example data point in black, with its observed uncertainties, plotted over our prior represented by a mixture of Gaussians in green.
We chose example points that evenly spanned the color space, but within each color bin, we chose a point at random.
The remaining panels show the posterior PDF in blue, the prior PDF in green, and likelihood of the data in black.
Most stars have posterior PDFs (blue) that are only slightly more precise than the data (black), which is well represented in this random sampling of points.
Examples 1, 3, and 4 are not very effected by the prior (green), where good data is good data.
Examples 2 and 5 show a more precise posterior PDF compared with the likelihood.
Example 5 looks very similar to the prior PDF and exemplifies that our prior tends to have the largest effect on stars in the red giant branch.}
\label{fig:posterior}
\end{figure}

Figure~\ref{fig:posterior} shows the various probability distributions of our inference for a few randomly-chosen objects: the likelihood (black), prior (green), and posterior PDF (blue) on the true parallaxes.
For the majority of the \gaia\ stars, our prior slightly increases the precision of the posterior PDF compared with the \gaia\ likelihood. Our parallax posterior PDFs are more precise than the \tgas\ catalog entries by a median factor of $1.2$. Examples of these slight increases in precision are visualized in Examples 1, 3, and 4 in Figure~\ref{fig:posterior} which show slightly narrower posterior PDFs compared to the likelihoods.
However, $14\%$ of our parallax posterior PDFs are more precise than the \tgas\ catalog by a factor of $>2$. This larger change in precision is visualized in Example 5, which shows a significantly narrower posterior PDF compared to the likelihood.
The nature of our prior is also visible: it is apparent that it is made of a mixture of components, and that its impact, and shape, strongly depend on the location of each star in \cmd\ space.

\begin{figure}
\centering
  \includegraphics[width=\textwidth]{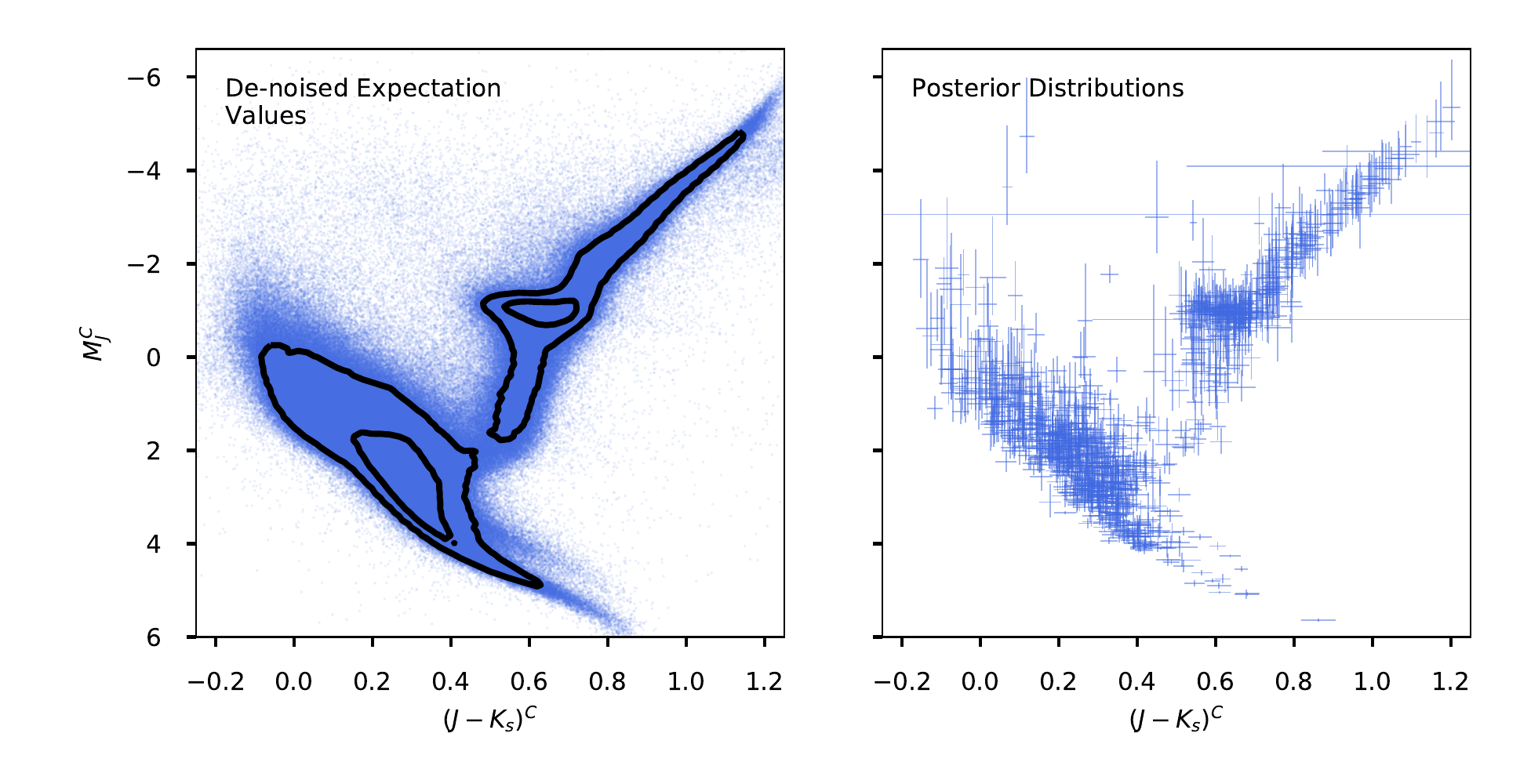}
\caption{Posterior Expectation Values: The distribution of expectation values of the parallax posterior PDFs transformed into an absolute magnitude vs color, the \cmd, of the full data set is shown in the left panel, with the $1\sigma$ and $2\sigma$ contours of the distribution shown in black.
The right panel shows a subsample of data set, plotting the expectation value and $1\sigma$ uncertainty of posterior PDF projected onto the \cmd.
Notice that some parts of this \cmd\ are tighter than the prior, or noise-deconvolved \cmd, shown in \figurename~\ref{fig:prior}.
The visual of the prior is a sampling of the Gaussians, but this is the expectation
  value, so it lies much closer to the center of the
  distribution. Similar to the toy model, where the maximum of the
  posterior PDF of the noisy points lies closer to the center of the true distribution than the true values.}
\label{fig:posteriorCMD}
\end{figure}

Figure~\ref{fig:posteriorCMD} shows the distribution of the expectation values of the posterior parallax PDFs projected back onto the \cmd\ space.
The left panel shows the distribution for all 1.4 million stars in our dataset, and the right panel shows a subsample of the dataset and includes the $1\sigma$ uncertainties on the posterior PDFs.
Compared with the raw data shown in Figure~\ref{fig:data}, it is clear the precision of the posterior PDFs is greater, especially for the red giant branch stars.

\begin{figure}
\centering
\includegraphics[width=\textwidth]{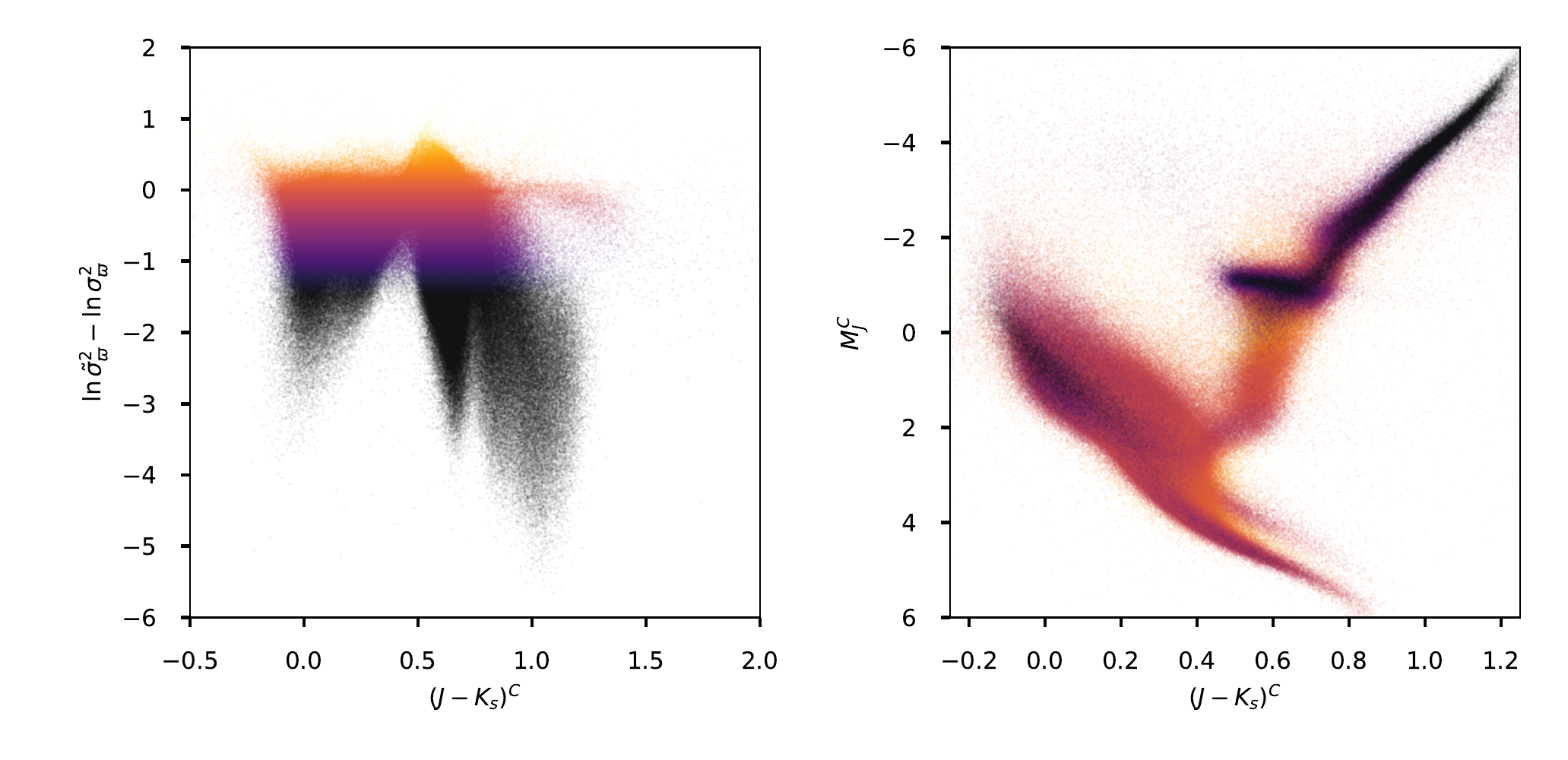}
\caption{Fractional Change in Variance: The left panel shows the natural log of the fractional change in the variance of the posterior PDF relative to the likelihood as a function of the stellar color. The coloring of the points in both panels is the fractional change in variance (the y-axis in the left panel) to guide the interpretation of the right panel. The median fractional change is 1.2. Some posteriors have significantly smaller variances shown in black, but some increase in variance, shown as the yellow bump at $(J-K_s)^{\corr}$ of about 0.6. The right panel shows the posterior point estimates on the CMD colored by fraction change in variance. There is significant changes for the red giant branch stars, as well as some main sequence stars.}
\label{fig:delta}
\end{figure}

This change in precision is further illustrated in Figure~\ref{fig:delta}, where we show the fractional changes in the variance. The left panel shows the natural log of the fractional change in variance as a function of $(J-K_s)^C$ color, with the $1\sigma$ and $2\sigma$ contours of the distribution over plotted.
The points are colored by the natural log of their fractional change in variance (the $y$ axis of the left panel) in both panels to help guide your eye for the right panel. The regions of \cmd\ space with the greatest improvements are shown in black, and the regions with the lowest improvements are shown in yellow.
The giants show obvious large improvements, as well as some main sequence stars.
Regions of large improvement are those with the largest photometric and parallax errors, which are the faintest observed objects.
This is a fairly standard result which we illustrated in the toy model above: the deconvolution of uncertainties is stronger in noisier regions of the data, leading to narrower features in the upper part of the \cmd, for example.
Stars with a color $(J-K_s)^C \sim 0.6$ show a decrease in precision due to the width of the prior there; the more vertical structure in the red giant branch of the \cmd.

\begin{figure}
\centering
\includegraphics[width=0.5\textwidth]{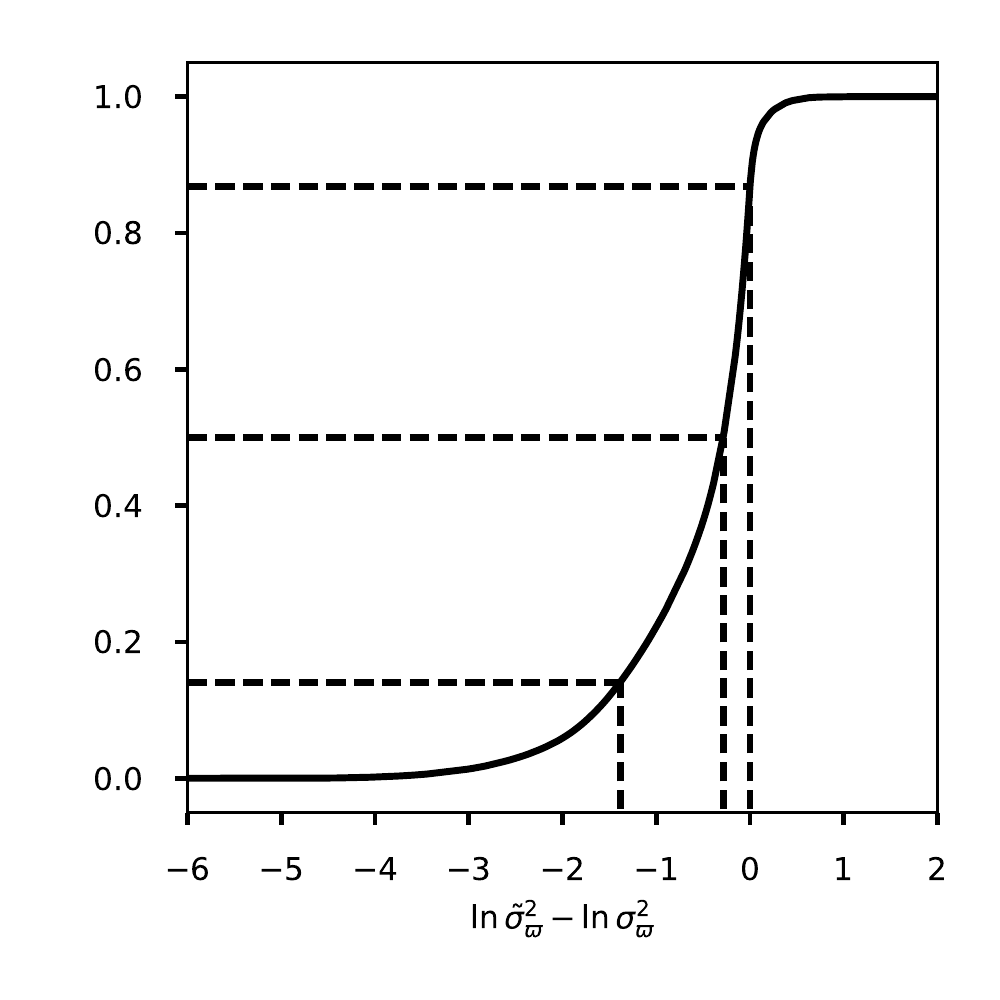}
\caption{Fractional Change in Variance: The cumulative distribution function of the natural log of the fractional variance change when comparing our posterior to the \tgas\ catalogue. Negative numbers represent stars which have a posterior that is narrower than the \tgas\ catalogue, which is true for a large fraction of the stars. We've drawn three dashed lines. One represents the median star $(y=0.5)$ which has a fractional variation of $x=-0.3=ln(1/1.2^2)$, corresponding to an increase in precision by a factor of 1.2. Another represents the fraction of stars $(y=0.14)$ which have a fractional variance greater than $x=-1.4=ln(1/2^2)$, corresponding to an increase in precision by a factor of 2. The final line, $x=0$, represents the fraction of stars with an increased precision, $83\%$.}
\label{fig:deltaCDF}
\end{figure}

Figure~\ref{fig:deltaCDF} shows the cumulative distribution function of the natural log of the fractional change in variance of the parallax posterior pdf relative to the \tgas\ catalog (the y-axis of Figure~\ref{fig:delta}). This more directly shows what was alluded to in Figure~\ref{fig:posterior}. For the majority of the \gaia\ stars, our prior slightly increases the precision of the posterior PDF compared with the \tgas\ variance. Our parallax posterior PDFs are more precise than the \tgas\ catalog entries by a median factor of $1.2$, shown as the dotted line at $y=0.5$.
The precision of our posterior PDFs is greater than or equal to the precision of the \tgas\ catalogue for $83\%$ of the stars, shown as the dotted line at $x=0$,
and $14\%$ of our parallax posterior PDFs are more precise than the \tgas\ catalog by a factor of $>2$, which corresponds to a fraction decrease in variance by a factor of $>4$, shown as the dotted line at $y=0.14$.

\begin{figure}
\centering
\includegraphics[width=0.5\textwidth]{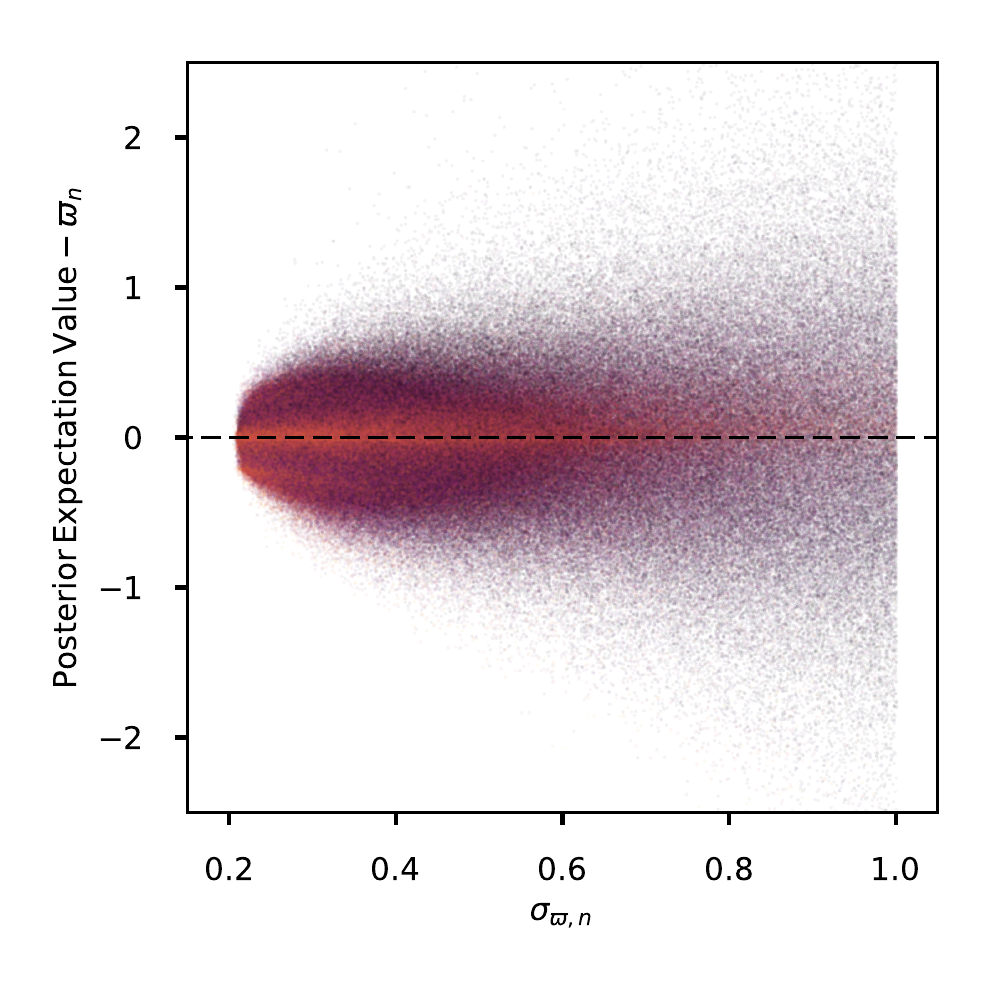}
\caption{Minimal bias introduced: The difference between the expectation value of our parallax posteriors and the \tgas\ parallax measurement as a function of the \tgas\ uncertainty. The points are colored by the fractional change in the variance of the posterior PDF compared with the \tgas\ likelihood, the same as Figure \ref{fig:delta}. The points are evenly distributed around $y=0$ showing that there is little bias introduced by inferred parallax PDFs. The distribution gets wider with larger \tgas\ uncertainties due to those lower signal to noise measurements being more influenced by the prior. They also tend to have the largest fractional change in their variance (shown as darker points)}
\label{fig:deltaParallax}
\end{figure}

We have shown that the posterior parallax estimates are more precise
than the likelihoods, both on average and especially for low
signal-to-noise stars.
It is also interesting to ask whether our data-driven prior leads to
biased inferences.
In Figure~\ref{fig:deltaParallax}, we show the shifts of the parallax
estimates from likelihood \tgas\ catalog entry to posterior
expectation.
This shows that while many parallax estimates change substantially,
they do not change on average, even at low signal-to-noise ratios.
Our conclusion is that the prior developed here does not introduce additional
bias into the \tgas\ parallax measurements.

\subsection{Data products}

The method produces a mixture-of-Gaussian parallax posterior PDF for every
star in the intersection of \tgas, \tmass, and the dust map.
Associated with this paper is an electronic table containing our results\footnote{\url{http://voms.simonsfoundation.org:50013/8kM7XXPCJleK2M02B9E7YIYmvu5l2rh/ServedFiles/}}, and everything
that is necessary to use them responsibly.
It includes the \tgas\ identifier, sky position, and dust-corrected
photometry.
It contains posterior expectations and variances for the stellar parallax.
It also contains the full posterior PDF for each star, in the form of a set of PDF values,
evaluated on a common grid in parallax.
Because all of the data are public, and all of the code used for this project is available
publicly under an open-source license, power users can generate any of the figures in this
paper or any other data outputs straightforwardly.

\subsection{Distances to M67}
As a test of the accuracy of these photometric parallax inferences, we inferred the parallaxes to stars in the open cluster M67.
M67 is at a Heliocentric distance of $800$--$900~{\rm pc}$ with an estimated age of $3$--$5~{\rm Gyr}$.
Previous work \citep{cbj15} has noted that distance estimates directly derived from the inverse of the observed (noisy) parallax are unreliable (significantly biased and uncertain) when the parallax signal-to-noise (SN) is below 5, $\varpi / \sigma_\varpi < 5$.

With \tgas\ parallaxes having a minimum uncertainty of about $0.3~{\rm mas}$, the SN reaches 5 for the most certain parallax measurements at about $700~{\rm pc}$, just before the distance to this cluster.
M67 is therefore a great test case for comparing parallax inferences with or without a prior.

\begin{figure}
\centering
  \includegraphics[width=\textwidth]{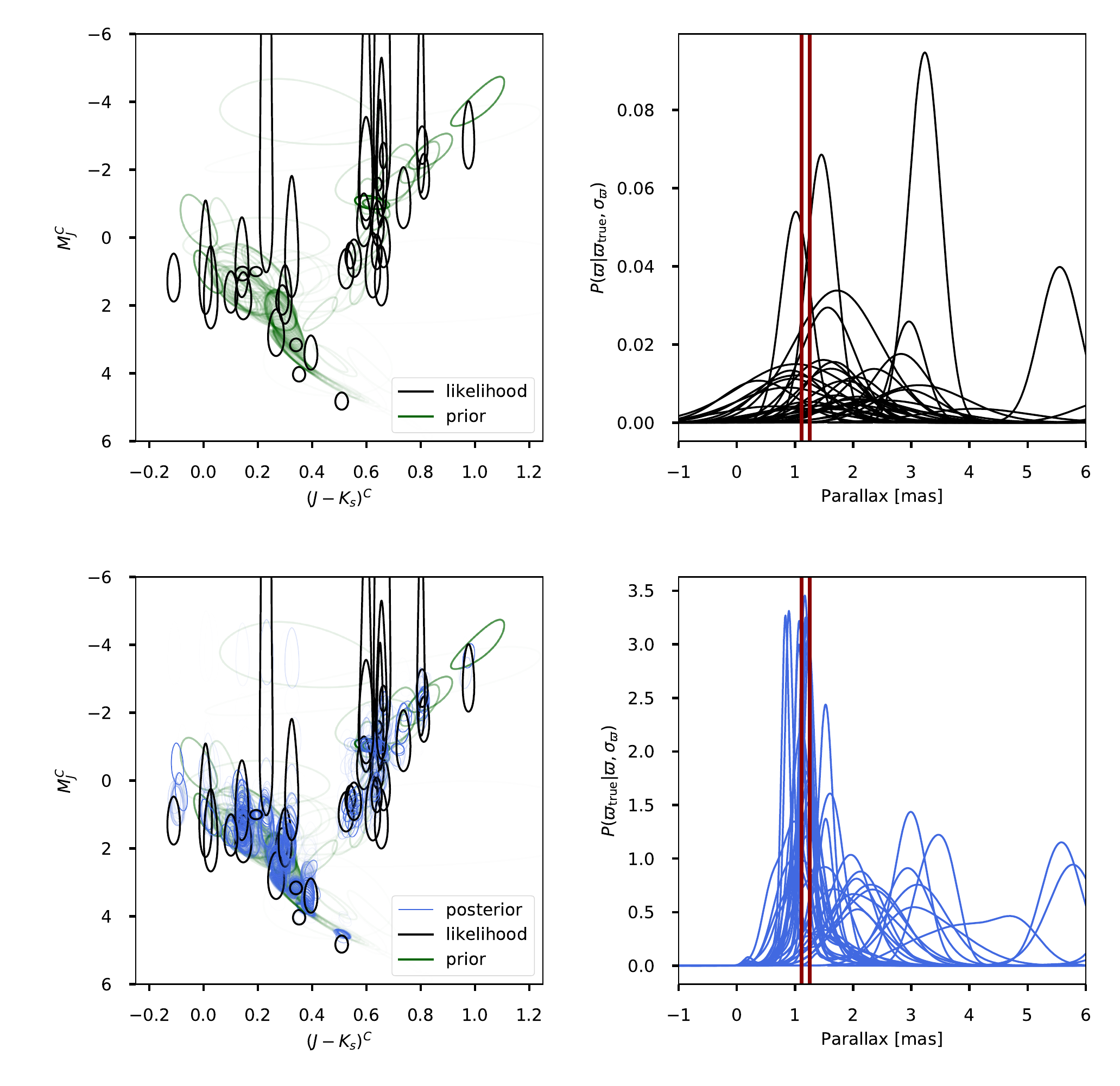}
\caption{Proof of concept: parallax inference to the open cluster M67.
The upper left panel visualizes the \gaia\ + \tmass\ likelihoods of the data, shown in black and projected onto the \cmd, for all the stars within our $1~{\rm deg}^2$ window centered on the position of M67.
Our prior, visualized as a mixture of Gaussians, is shown in green.
The upper right panel shows the \tgas\ parallax likelihoods in black, with the ``true" distance to M67, determined by previous methods, bracketed by the red lines.
Similar to the upper left panel, the lower left panel shows the likelihoods of the data in black and our prior in green, as well as the posterior PDFs in blue.
The posterior PDFs are a mixture of Gaussians as well, each a product of the likelihood with a Gaussian in our prior.
The lower right panel shows the projection of the 2D posterior PDF onto parallax space.
The posterior PDFs are much more narrow and clustered than the \tgas\ likelihoods, showing that our prior generates posterior PDFs that are accurate and more precise than \gaia\ alone.}
\label{fig:m67}
\end{figure}

We select sources from our catalog within a $1~{\rm deg}^2$ window centered on the position of M67; the cluster is visible as an over-density of stars in the sky positions alone.
To be clear, we do no prior selection on previously known members of M67, but include all stars observed by \tgas\ within this window on the sky.
Keep in mind, the (low) number of M67 members seen here is constrained by the (unknown) selection effects of the \tgas\ dataset.
With our converged dust values, and its associated prior, we calculate the posterior over $\varpi_{\true}$ for each star within this window on the sky around M67.
\figurename~\ref{fig:m67} shows the comparison of the likelihoods and posteriors.
The vertical red lines bracket the previously measured parallaxes of, or distances to, the cluster.
The likelihoods of the observed parallaxes, shown in the top row in black, are broad and the cluster is not at all obvious.
The posterior PDFs, shown in the bottom row in blue, are more sharply peaked showing an increase in precision.
The posterior PDFs also have modes that are more similar to the previously determined distance to the cluster (vertical lines) for far more stars.
This shows that our prior is not only making parallaxes more precise but also (possibly) increasing the accuracy.

\section{Discussion}

We have shown that it is possible to obtain photometric parallaxes for
distant stars in the \gaia\ \tgas\ catalog without any use of physical stellar
models, nor stellar density models of the Milky Way.
We used the geometric parallaxes to calibrate a photometric
model that is purely statistical, which is a model of the data rather than
a model of stars \foreign{per se}.
This opens up the possibility of completing the goals of the \gaia\ Mission
without building in unnecessary assumptions about the mechanical properties
of stars or the Galaxy.

We obtained the photometric parallaxes in this project
by building a data-driven model of \cmd\ of the stars in the \tgas\ data set,
and using it as a prior PDF for Bayesian inference.
The posterior PDFs for distance that we obtain are, in general, much
narrower than the likelihood functions delivered by the
\gaia\ Mission, and therefore the distance estimates (or,
equivalently, parallax estimates) are much more precise.
It is not surprising that a Bayesian inference provides more
precise inferences than the likelihood function alone; Bayesian
inferences bring in new information that decrease variance (but
can introduce bias).
We have shown that, in addition to pure precision improvements, at
least some aspects of \emph{accuracy} have been improved as well:
The posterior distance estimates to stars in (at least one) stellar
cluster are much more clustered than likelihood-based distance
estimates, and they cluster around a sensible value.
Accuracy is demonstrated by this test because each star in the cluster is
subject to a different, unique distance prior (see \figurename~\ref{fig:posterior})
and yet the posterior PDFs are all pulled to the same parallax value.

Although we have not performed principled hierarchical Bayesian inference in this
project, we built our prior for each star's parallax by
performing a statistically responsible deconvolution
of the \gaia\ \tgas\ data that is justifiable under a clear set of
assumptions.
This deconvolution, using a reasonable noise model and an
assumption of stationarity, shrinks the parallax uncertainties, most dramatically for the stars measured at lower signal-to-noise.
Because the method we use, \xd, which is
an Empirical Bayesian maximum-marginalized-likelihood estimator, accounts for
heteroskedastic noise, we are able to build our photometric parallax
model---which is a model of the \cmd---using
all the data, not just the data with the highest signal-to-noise.
So, the generated model for the \cmd\ is representative for
our subset of the \tgas\ Catalog.
Because our model is representitive of our full subset catalog, we expect any (sensible) parallax estimates
we generate from our parallax posterior PDFs to be only weakly biased.

\subsection{Comparison with other priors}
\begin{figure}
\centering
  \includegraphics[width=\textwidth]{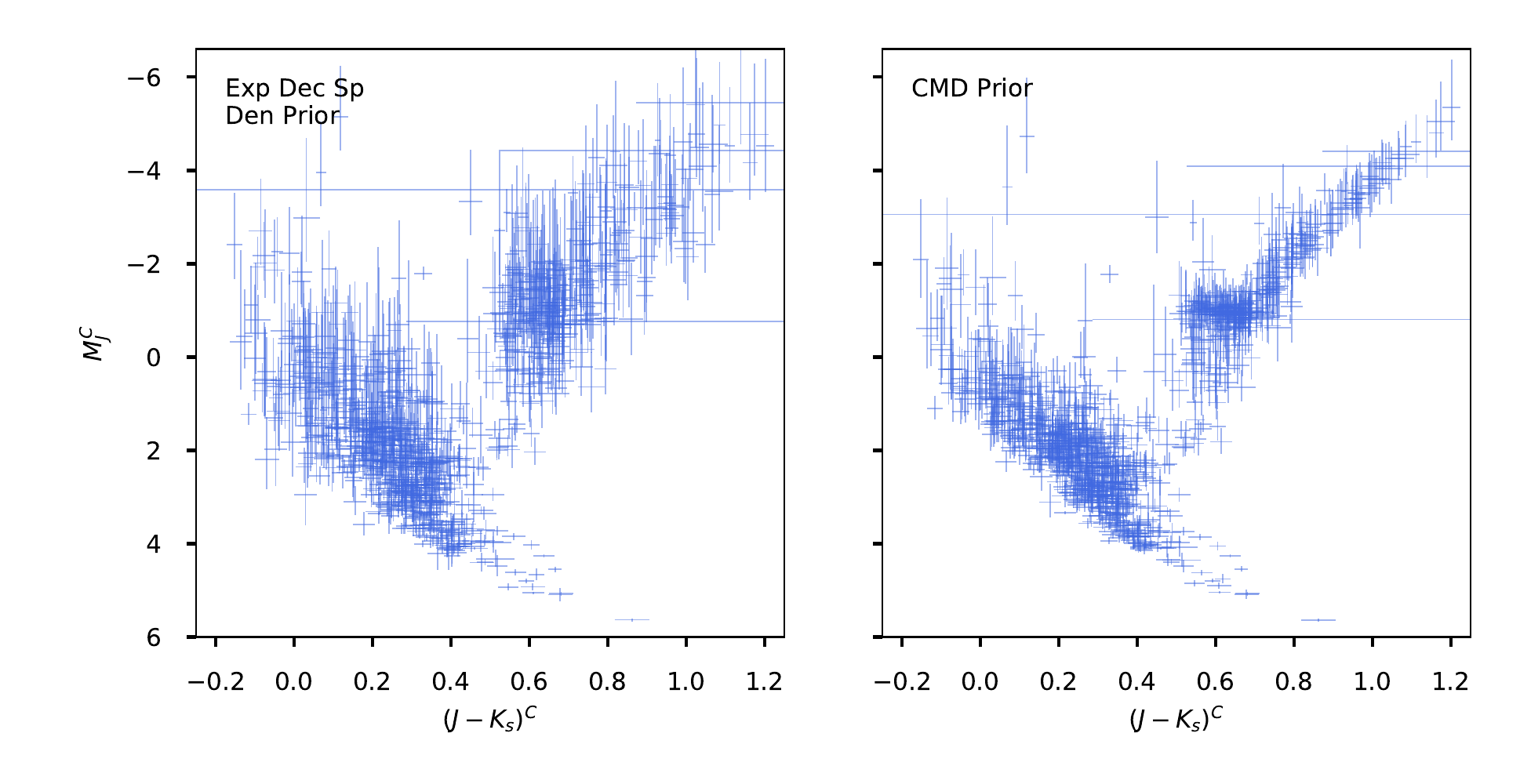}
\caption{Comparison with a common spatial prior: A visualization of the posterior PDF for a subsection of \tgas\ stars, shown as the expectation value and $1\sigma$ uncertainty. The left panel shows the posterior PDF using the exponentially declining spatial density, a common space-density prior used when inferring distances. The right panel shows the posterior PDF using our \cmd\ prior. The scale of the exponentially declining spatial density prior is $1.35~{\rm kpc}$, the optimal value found in \cite{astraatmadja16a}. Using the \cmd\ as a prior generates a posterior PDF that has smaller variance.}
\label{fig:comparePrior}
\end{figure}

There are other options for prior information to include when inferring distances to stars.
One option is the exponentially declining spatial density (EDSD) of stars \citep{astraatmadja16a}.
It is parameterized by a scale length, and has the nice property that it is smooth out to very large distances so that the posteriors are also very smooth.
It is a fairly weak prior.
\figurename~\ref{fig:comparePrior} shows a comparison of the EDSD prior with our \cmd\ prior.
Here the parallax posterior PDF expectation value, and $1\sigma$ uncertainty, for a subset of stars in our dataset is converted into an absolute magnitude and visualized as the \cmd.
The left panel visualizes results using the EDSD prior, and the right panel visualizes results using our \cmd\ prior.
The variance for the \cmd\ prior is significantly smaller than the variance for the EDSD prior.
In general, EDSD is a fairly broad prior so the median \gaia\ star has a posterior variance equal to the likelihood variance.
EDSD has the largest effect on the most noisy and negative parallaxes, and has the nice property of bringing these noisy or negative measurements to reasonable parallaxes.

\subsection{What's That Feature?}
\begin{figure}
\centering
  \includegraphics[width=\textwidth]{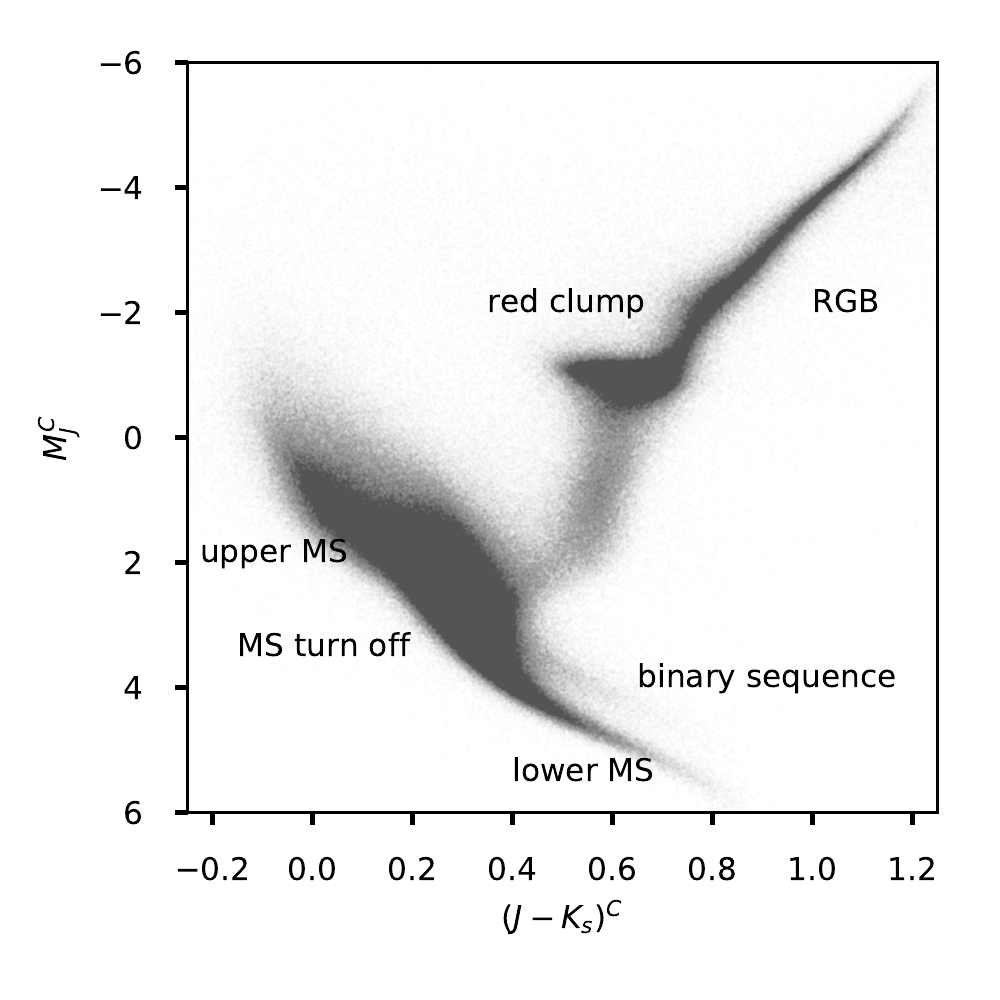}
\caption{Features in the \cmd: The expectation values from the posterior PDFs of the true parallax projected onto the \cmd.
Here we have noted the features previously found in the \cmd\ that are also highlighted nicely by the \gaia\ + \tmass\ data.
The upper and lower main sequence, as well as the main sequence turn off are easily seen.
There is a plausible binary sequence of stars, as well as a prominent red clump, and straight and narrow red giant branch. }
\label{fig:wtf}
\end{figure}

The denoised expectation values of the parallax projected into the \cmd\ show many interesting and familiar features; we visualize this \cmd\ again in \figurename~\ref{fig:wtf} and highlight some of these features.

The lower main sequence is very narrow while the upper main sequence
is much wider. This may be an age and metallicity variation on the main sequence mixed
with a Malmquist bias of seeing a much larger volume of upper main
sequence stars \citep{malmquist22}. There is a plausible binary sequence of stars with
about twice the brightness as the main sequence. The helium burning
red clump is prominent for the volume DR1 probed, and the red giant
branch is fairly straight and narrow. The main sequence turn off is
noticeable and also surprisingly narrow, possibly a reflection of the
star formation history of the local volume also mixed with a Malmquist
bias.

\subsection{Critical discussion of assumptions}

In Section~\ref{sec:method} we listed a set of assumptions upon which
the photometric parallax method in this project is based.
Here we return to these assumptions, and discuss them critically in more
detail, under the same set of assumption labels:
\begin{description}
\item[stationarity]
  We have assumed that (locally) the morphology of
  the \cmd\ is independent of signal-to-noise; that is, we have assumed that
  both the low and high
  signal-to-noise objects are being drawn from the same distribution.
  This isn't precisely true because nearby stars in \tgas\ are
  mostly main sequence stars, while we can see luminous giants to much larger
  distances. So the red giant branch is a population of
  relatively lower signal-to-noise objects compared with the main
  sequence. This does not violate the stationarity assumption in
  detail, because the low and high signal-to-noise objects have to be
  drawn from the same distribution within some feature or patch of the
  \cmd\ space. Thus, this has a small effect on our results, and could be improved by modeling the noise distribution simultaneously with the \cmd.
\item[selection] The biggest limitation of our method for generating
  the \cmd\ from the data is that we have not used a
  \tgas\ selection function, completeness estimate, or
  inverse-selection-volume corrections.  This model is a model of the
  \tgas\ Catalog (or really the \tgas--\tmass--\psone\ intersection), not of
  the stars in the Milky Way, nor any volume-limited subsample of the
  Milky Way.  Importantly, because we used \emph{all} of the stars in
  the \tgas--\tmass--\psone\ intersection, the model \emph{is} representative of
  the full intersection, even the distant parts, where all the
  parallaxes are measured at low signal-to-noise.  This is in contrast
  to techniques that might build the model from only the high
  signal-to-noise sources; such a model would be biased towards
  stellar types and compositions found locally, as well as
  lower-luminosity stars, and stars which happened to get good
  parallaxes.  However, our use of the entire
  \tgas--\tmass--\psone\ intersection without any selection function restricts
  the use of our \cmd\ model as a prior to inferences within that same
  \tgas--\tmass--\psone\ intersection.  It would be biasing to use this same
  prior to infer distances for the full billion-star catalog released
  in \gaia\ DR1 alongside \tgas, or for stars fainter or brighter or
  bluer than those included in the \tgas--\tmass--\psone\ intersection.
  A similar argument can be made for \textit{internal} selection effects, such as inhomogeneities of the depth and noise distributions, which we have neglected.
  Those inhomogeneities are unlikely to affect the \cmd\ results; we effectively model the average distribution of the \tgas--\tmass--\psone\ intersection.
  By modeling the selection function in detail, one could improve the quality of the inferred \cmd, its effect on individual objects, and its connection to physical models of the Milky Way.
\item[big data] The empirical Bayesian methodology is a good
  approximation to full hierarchical inference in the limit of large
  numbers of data points, such that no individual star is carrying a
  lot of weight in the model of the \cmd.  With its millions of stars,
  the \tgas\ Catalog appears to safely live in this limit.  However,
  because there are parts of the \cmd\ that are poorly populated, even
  in \tgas, the empirical Bayes approximation may be bad for some
  portions of the \cmd.  In particular, there appear to be essentially
  no white dwarf stars, and other classes (like bright red giant
  branch stars) have few to no exemplars in the data set at high
  signal-to-noise (in parallax).  For all these reasons, we do think
  that this approximation is not perfectly safe, and it is a goal to
  explore full hierarchical modeling in the coming years (and see, for
  example, \citealt{leistedtHogg2017}).
\item[noise model] Everything in both the establishment of the data-driven
  prior and the use of it for parallax inference has been done under the deep
  assumption that the \gaia\ noise model is correct, and an additional
  assumption that the parallax uncertainty is always significantly larger
  than the photometric uncertainty in the $J$ band, even after dust correction.
  In the unlikely event (given these choices) that the
  uncertainties are over-estimated, we could be over-deconvolving.
  That is, the data-driven model could be producing a tighter or more
  informative distribution than what would truly be observed in a much
  higher signal-to-noise experiment or survey. That is, incorrect uncertainty
  estimates will bias the empirical-Bayes prior we obtain. On the inference
  side, wrong uncertainty estimates lead to further biases in the derived photometric parallaxes.
  Since we are explicitly ignoring the (non-Gaussian, in parallax space) photometric
  and dust-extinction uncertainties, our noise estimates are certainly biased, and
  this will translate into small biases in posterior parallax estimates.
\item[dust] Within our probabilistic framework, we improperly handle
  dust. We are taking a point estimate of the probabilistic dust model
  at a quantile of our distance posterior, correcting our photometry
  using this point estimate, and then re-deriving the prior. A full,
  proper account would sample the dust model at distances sampled from
  the posterior, and propagate those samples back into the prior
  during the iterative process. The reason we don't do a full
  probabilistic treatment, is that this proper dust distribution would
  not be a Gaussian distribution, which \xd\ requires. We could
  calculate this more complex dust distribution and estimate it as a
  Gaussian distribution to feed to \xd, however, this is beyond the
  scope of this paper.
\item[mixture of Gaussians] The \cmd\ model is a mixture of Gaussians,
  and furthermore fixed at $K=128$ components. This setting was chosen
  after some experimentation as being able to capture features but still
  easily optimized. We performed some experiments
  that suggest that larger $K$ would lead to better models, even in the
  conservative sense of a cross-validation score. Choosing an optimized or
  optimal number for this mixture is a natural extension of this work.

  Beyond this, the choice of the Gaussian form itself for the mixture is
  questionable. The Gaussian has great properties, that we capitalize on
  both at the prior-generation stage (with \xd), and at the photometric-parallax
  stage (capitalizing on Gaussian product rules). We could move to other
  forms, but they would be extremely expensive, computationally.
\item[no physics] Although it is presented as an advantage of our
  method that we never make any use of physical models, it is also
  brings some significant disadvantages: For example, the \cmd\ contains
  features that are obvious products of stellar evolution, but doesn't in
  itself inform the theory or models of stellar evolution, because our
  mixture-of-Gaussian \cmd\ model is divorced from any theoretical ideas
  about stars. For another, stellar evolution models are in fact very
  predictive and successful; our model does not capitalize in any way on
  those successes. A next-generation model might be designed to be data-driven
  but nonetheless capitalize on physics-driven successes. One framework
  for that would be to fit not the \cmd\ but rather deviations of the \cmd\ away
  from a good physics-based model prediction. This would also require modeling the selection function of the data under consideration.
\end{description}

The outputs of this project include posterior PDFs for stellar parallaxes.
It is tempting to treat these outputs as equivalent to some kind of catalog of
measurements, as we treated the original parallax measurements from \tgas.
That is, it is tempting to treat these as simply ``better'' measurements.
Taken one star at a time, this is permitted and true, and the basis for our
claim that we have de-noised the \tgas\ Catalog.
However, there are important differences, as there are with all
probabilistic catalogs (\citealt{hogg11}; \citealt{portillo17}), between the
original \tgas\ measurements and the de-noised measurements.
For users of our output, the most important difference between our
output and the \tgas\ data input is that the \tgas\ data is the representation
of a likelihood function, whereas our ouptut is a
representation of a posterior PDF, one per star.
Posterior information used naively can lead to serious statistical
errors, because each data point has had a prior multiplied in.
Multiplicative uses of the data will effectively take the prior
PDF to a large power.
We have (in previous work)
given examples of correct uses of posterior samplings or PDFs
for subsequent inferences (\citealt{hogg08}; \citealt{dfm14}); we
encourage power users to consult those methodological
contributions before engaging with these outputs.

\section{Conclusion}
We forgo an explicit conclusion; we attempt to summarize the conclusions in the Abstract above.

\acknowledgments It is a pleasure to thank
  Ana Bonaca (Harvard),
  Andy Casey (Monash),
  Stephen Feeney (Flatiron),
  Dustin Lang (Toronto),
  David Spergel (Flatiron),
and the attendees at the Stars Group Meeting at the Flatiron Institute
Center for Computational Astrophysics for comments and input.
This project was developed in part at the 2016 \acronym{NYC} \gaia\ Sprint, hosted
by the Center for Computational Astrophysics at the Simons Foundation
in New York City.

This work has made use of data from the European Space Agency (\acronym{ESA})
mission \gaia\footnote{\url{http://www.cosmos.esa.int/gaia}}, processed by the \gaia\ Data
Processing and Analysis Consortium\footnote{\url{http://www.cosmos.esa.int/web/gaia/dpac/consortium}} (\acronym{DPAC}). Funding for the
\acronym{DPAC} has been provided by national institutions, in particular the
institutions participating in the \gaia\ Multilateral Agreement.

This publication makes use of data products from the Two Micron All
Sky Survey, which is a joint project of the
University of Massachusetts and the Infrared Processing and Analysis
Center/California Institute of Technology, funded by the National
Aeronautics and Space Administration and the National Science
Foundation.

This research was partially supported by
  the \acronym{NSF} (\acronym{AST-1517237}),
  \acronym{NASA} (grant \acronym{NNX12AI50G}),
  and the Moore-Sloan Data Science Environment at \acronym{NYU}.
  BL was supported by the National Aeronautics and Space Administration through Einstein Postdoctoral Fellowship Award Number PF6-170154.
It made use of the \acronym{NASA} Astrophysics Data System.
All the code used in this project is available online\footnote{\url{https://github.com/andersdot/photoParallax}} under an open-source license.

\software{
The code used in this project is available from
\url{https://github.com/andersdot/photoParallax}.
This research utilized the following open-source \textsl{Python} packages:
    \textsl{Astropy} (\citealt{Astropy-Collaboration:2013}),
    \textsl{matplotlib} (\citealt{Hunter:2007}), and
    \textsl{numpy} (\citealt{Van-der-Walt:2011}).
This work additionally used the \gaia\ science archive
(\url{https://gea.esac.esa.int/archive/}).
}

\bibliography{gaia}

\end{document}